\documentclass[pdflatex,sn-nature,twocolumn]{sn-jnl}
\usepackage{geometry}
\usepackage{lineno}
\usepackage{graphicx}  
\usepackage{dcolumn}   
\usepackage{bm}        
\usepackage{amssymb}   
\usepackage{amsmath,stmaryrd}
\usepackage{blkarray, multirow, graphicx, diagbox, color, colortbl}
\usepackage[dvipsnames]{xcolor}
\usepackage{bbm, bbold}
\usepackage{glossaries}
\usepackage{algorithm}
\usepackage{algpseudocode}
\usepackage{hyphenat}
\usepackage{ifthen}
\usepackage{dsfont}
\usepackage{xkeyval}
\usepackage{moreverb}
\usepackage{rotating}
\usepackage{wrapfig}
\usepackage{slashbox}
\usepackage{xurl}
\usepackage{xspace}
\usepackage{nicefrac}
\usepackage[]{units}
\usepackage{physics}
\usepackage{booktabs}
\usepackage{mathtools}
\usepackage{tikz}
\usetikzlibrary{arrows.meta}
\usepackage{braket}
\usepackage[inline]{enumitem}
\usepackage{tabto}
\usepackage{listings}
\usepackage[normalem]{ulem}
\usepackage{xstring}
\def\ReplaceStr#1{%
	\IfSubStr{#1}{p}{%
		\StrSubstitute{#1}{p}{.}}{#1}}
\usepackage[english]{babel}

\usepackage[caption=false]{subfig}
\usepackage[capitalise]{cleveref} %

\usepackage{cuted}
\usepackage{chemformula}
\usepackage{algorithm}

\newcommand{\updated}[1]{\textcolor{black}{#1}} 

\graphicspath{{figures_paper/}}

\begin{document}

\def\thetitle{Digitally Optimized Initializations for Fast Thermodynamic Computing}
\title[Digitally Optimized Initializations]{\thetitle}

\author*[1]{\fnm{Mattia} \sur{Moroder}}\email{moroderm@tcd.ie}
\author[1]{\fnm{Felix C.} \sur{Binder}}
\author[1]{\fnm{John} \sur{Goold}}

\affil[1]{\orgdiv{School of Physics}, \orgname{Trinity College Dublin},
  \orgaddress{\street{College Green}, \city{Dublin 2},
  \postcode{D02K8N4}, \country{Ireland}}}

\abstract{Thermodynamic computing harnesses the relaxation dynamics of physical systems to perform matrix operations. A key limitation of such approaches is the often long thermalization time required for the system to approach equilibrium with sufficient accuracy. Here, we introduce a hybrid digital-thermodynamic algorithm that substantially accelerates relaxation through optimized initializations inspired by the Mpemba effect. In the proposed scheme, a \updated{standard} digital processor efficiently computes an initialization that suppresses slow relaxation modes, after which the physical system performs the remaining computation through its intrinsic relaxation dynamics. We focus on overdamped Langevin dynamics for quadratic energy landscapes, analyzing the spectral structure of the associated Fokker-Planck operator and identifying the corresponding optimal initial covariances. This yields a predictable reduction in thermalization time, determined by the spectrum of the encoded matrix. We derive analytic expressions for the resulting speedups and numerically analyze thermodynamic implementations of matrix inversion and determinant computation as concrete examples. Our results show that optimized initialization protocols provide a simple and broadly applicable route to accelerating thermodynamic computations.}

\maketitle
\section*{Introduction}
Thermodynamic computing is an emerging analog computational paradigm in which linear algebra problems are solved by exploiting the relaxation dynamics of physical systems~\cite{Conte2019,Wolpert2024,Aifer2024}. 
This approach has attracted significant interest mainly for two reasons.
First, thermodynamic computing naturally implements probabilistic computation, making it a promising platform for probabilistic inference, including Bayesian machine-learning algorithms, which are potentially far more energy-efficient on stochastic hardware than on deterministic digital devices~\cite{Melanson2023,jelincic2025,Rolandi2026}.
Second, it has been shown that several central linear algebra operations can exhibit favorable scaling on thermodynamic hardware compared to standard digital architectures~\cite{Aifer2024, Bartosik2024, Duffield2025}.
More recently, thermodynamic hardware has also been proposed as a platform 
for generative modeling~\cite{Whitelam2026}, further broadening the scope of thermodynamic computing.

In parallel, in the field of nonequilibrium thermodynamics there has been intense research activity on anomalous thermalization phenomena, most notably the Mpemba effect~\cite{Lu2017}, whereby systems prepared further from equilibrium can relax faster than systems initially closer to equilibrium.
Originally observed in classical settings~\cite{Klich2019, Kumar2020, Gal2020,Kumar2022, Teza2023, Teza2025}, the Mpemba effect and related phenomena have since been extensively studied also in quantum systems~\cite{Nava2019, Carollo2021, Bao2022, Kochsiek2022, Ivander2023, Wang2024, Moroder2024, Strachan2024, Xu2025, Longhi2025,Ares2023Nat, Ares2025, Rylands2024, Turkeshi2024, Liu2024, Li2025, Summer2026, Beato2026}. 
At a qualitative level, anomalously fast equilibration can arise when the initial state has little or no overlap with the slowest decaying modes of the dynamics, thereby bypassing the dominant relaxation bottlenecks~\cite{Lu2017, Carollo2021}.

In this work, we show that the Mpemba effect can be harnessed to accelerate key stages of thermodynamic computing routines.
We propose a hybrid digital-thermodynamic protocol in which optimized initial conditions are computed on a conventional digital processor and then encoded into the thermodynamic hardware, as summarized pictorially in Fig.~1.
A continuous-variable implementation of thermodynamic computing based on coupled harmonic oscillators~\cite{Melanson2023} is considered.
As concrete examples of single-stage and multi-stage algorithms, we analyze matrix inversion and matrix-determinant computation and quantify the resulting Mpemba-induced speedups for different classes of random matrices.

\begin{figure}[t] 
    \centering
    \subfloat[\label{subfig:first:a}]{
        \includegraphics[width=0.95\columnwidth]{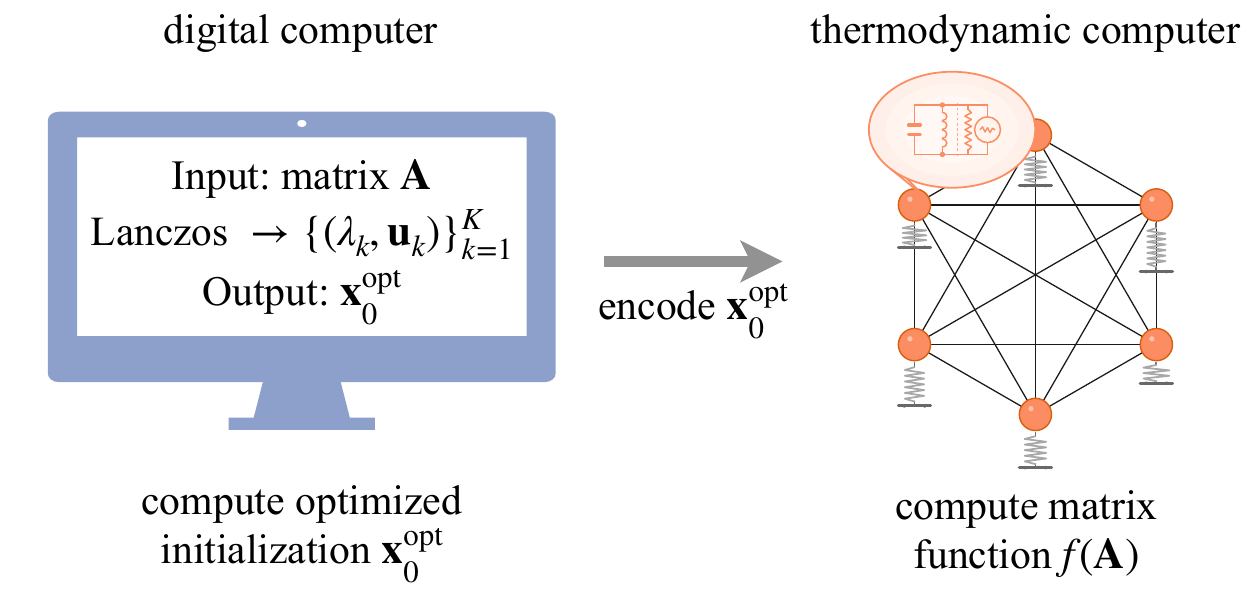}
    }
    \vspace{-0.4cm} 
    \subfloat[\label{subfig:first:b}]{
        \includegraphics[width=0.95\columnwidth]{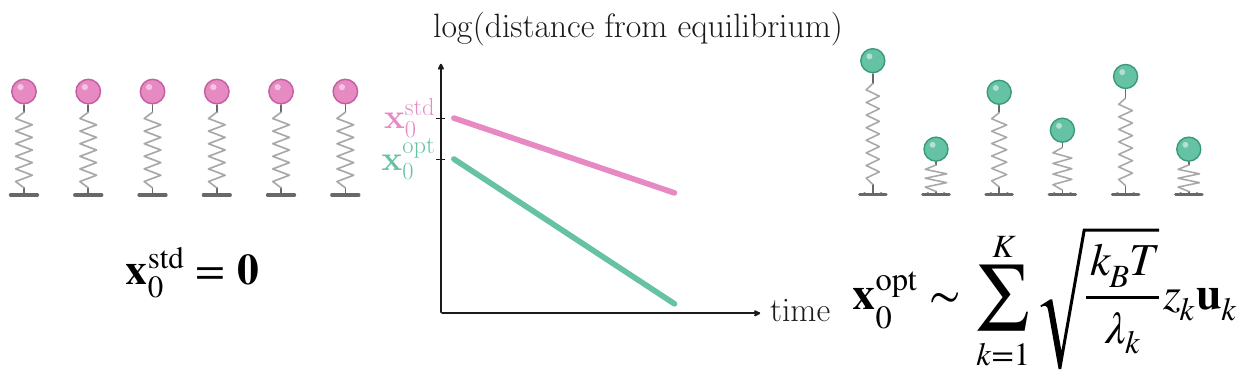}
    \hspace{-0.1cm}
    }
    \caption{\textbf{Hybrid digital-thermodynamic computing protocol with Mpemba-inspired initialization.}
(a) A \updated{digital} processor computes an optimized initialization for the thermodynamic computer. Given an input matrix $\textbf{A}$, a Lanczos routine efficiently extracts the $K$ smallest eigenpairs, which are used to prepare an initial condition $\mathbf{x}_0^{\mathrm{opt}}$. This initialization is encoded in a thermodynamic device consisting of coupled harmonic oscillators (implemented as LC electrical circuits) that relax toward equilibrium, from which various matrix functions $f(\mathbf{A})$ can be estimated. 
(b) Comparison between standard and optimized initialization. Starting from the trivial initial state $\mathbf{x}_0^{\mathrm{std}}=\mathbf{0}$ generally excites all relaxation modes, whereas the optimized initialization prethermalizes the $K$ slowest modes, leading to a faster thermalization following a Mpemba-type speedup.}
\label{fig:first}
\end{figure}

\section*{\updated{Theory and Methods}}
\subsection*{\updated{Thermodynamic computing in a nutshell}}
In thermodynamic computing, matrix operations are mapped to the thermalization dynamics of physical systems~\cite{Conte2019}.
A particularly powerful realization of this idea encodes the problem in the potential energy of a collection of coupled harmonic oscillators.
The system is allowed to thermalize, and the solution is read out from equilibrium averages, fluctuations~\cite{Aifer2024} or correlations~\cite{Duffield2025}.
Specifically, a $d\times d$ symmetric positive definite matrix $\mathbf{A}$ can be encoded as the all-to-all coupling of $d$ harmonic oscillators
\begin{equation}
V(\mathbf{x}) = \frac{1}{2}\sum_{i,j} A_{ij} x_i x_j ,
\label{eq:quadratic:potential}
\end{equation}
where $x_i$ denotes the displacement of the $i$-th oscillator.
At thermal equilibrium with temperature $T$, the system obeys the Boltzmann distribution
$p_\mathrm{eq}(\mathbf{x}) \propto \exp(-1/(k_BT) V(\mathbf{x}))$. For quadratic potentials such as~\cref{eq:quadratic:potential}, $p_\mathrm{eq}$ is a zero-mean multivariate Gaussian with covariance
\begin{equation}
\langle x_i x_j \rangle = k_B T\, (\mathbf{A}^{-1})_{ij},
\label{eq:equilibrium:covariance}
\end{equation}
where $\langle\cdot\rangle$ denotes expectation with respect to the
relevant probability distribution (here $p_{\mathrm{eq}}$).
As a result, the inverse of $\mathbf{A}$ can be obtained by sampling equilibrium fluctuations of the physical system.
More
generally, this mapping enables a range of thermodynamic encodings for matrix computations, including for instance the solution of linear systems, the computation of matrix determinants~\cite{Aifer2024}, and the evaluation of matrix exponentials~\cite{Duffield2025}.
In most cases, a substantial part of the computational cost is set by the thermalization dynamics of the underlying physical system, motivating the study of strategies to accelerate relaxation toward equilibrium.

Crucially, thermodynamic computing hardware has recently been realized experimentally~\cite{Melanson2023}.
In this platform, harmonic oscillators are represented by LC electrical circuits that are thermalized by externally injected current noise, whose intensity sets the effective temperature of the system.
The node voltage $v_i$ of the $i$-th circuit plays the role of the oscillator displacement $x_i$.
The matrix of interest $\mathbf{A}$ is encoded directly in the capacitive couplings between the LC circuits, thereby defining the quadratic interaction potential~\cref{eq:quadratic:potential}.
The computations proceed as follows.
First, the matrix $\mathbf{A}$ is programmed into the hardware via the capacitive coupling network.
The system is then initialized in a standard reference configuration, denoted $\mathbf{x}_0^{\mathrm{std}} = \mathbf{0}$, and allowed to thermalize.
After thermalization, a desired matrix function $f(\mathbf{A})$ is estimated from measured statistical observables.
Since the dynamics is stochastic, expectation values can be obtained in two equivalent ways:
Either by repeating the experiment multiple times and averaging over an ensemble of independent trajectories, or by exploiting ergodicity and performing a long-time average along a single trajectory.
Throughout this article, we focus on the ensemble-averaged formulation, which provides a convenient and transparent framework to analyze the impact of optimized initializations.

\subsection*{\updated{The Mpemba effect in a nutshell}}
\label{subsec:mpemba}
The observation that hot water can sometimes cool faster than cold water dates back to Aristotle~\cite{Aristotle}, and was rediscovered experimentally by Mpemba and Osborne in the 1960s~\cite{Mpemba1969}, lending the effect its modern name. While the microscopic mechanism underlying the freezing of water remains 
debated, Lu and Raz~\cite{Lu2017} showed that anomalous relaxation is a generic feature of systems coupled to a thermal bath: a state prepared further from equilibrium can relax faster than one starting closer to it, without any 
phase transition involved. They considered a system whose probability distribution $p(t)$ over its state space evolves according to a linear master equation
\begin{equation}
    \partial_tp= \mathcal{L} p,
    \label{eq:linear:diff:eq}
\end{equation}
where $\mathcal{L}$ is a time-independent generator satisfying detailed 
balance with respect to the equilibrium distribution $p_\text{eq}$. 
Detailed balance ensures that $\mathcal{L}$ has real, non-positive 
eigenvalues $-\gamma_n \leq 0$, ordered as $0 = \gamma_1 < \gamma_2 \leq 
\gamma_3 \leq \ldots$, where $\gamma_1 = 0$ corresponds to the stationary 
eigenfunction $p_\text{eq}$. The eigenfunctions $\{\phi_n\}$ form a 
complete basis, and the general solution can be written as
\begin{equation}
    p(t) = p_\text{eq} + \sum_{n \geq 2} c_n \, e^{-\gamma_n t} \, \phi_n,
\end{equation}
where the coefficients $c_n$ are fixed by the initial distribution $p(0)$. 
A natural measure of the distance from equilibrium is
\begin{equation}
    D(t) \equiv \| p(t) - p_\text{eq} \|,
\end{equation}
for some suitable norm $\|\cdot\|$. For a generic initial state with 
$c_2 \neq 0$, the long-time decay of $D(t)$ is controlled by the slowest 
decaying mode, $D(t) \sim |c_2| e^{-\gamma_2 t}$, and the thermalization 
timescale is set by $\gamma_2^{-1}$.

The Mpemba effect arises when two initial distributions $p_A(0)$ and 
$p_B(0)$ satisfy $D_A(0) > D_B(0)$ (that is, $p_A$ starts further from 
equilibrium) yet their distance curves cross at a finite time $t^*$, so 
that $D_A(t) < D_B(t)$ for all $t > t^*$. A sufficient condition for this 
anomalous crossing is that $c_2^A = 0$, i.e., the initial distribution 
$p_A$ is orthogonal to the slowest decaying eigenmode $\phi_2$. In this 
case, the asymptotic decay of $D_A(t)$ is governed by the faster rate 
$\gamma_3 > \gamma_2$, and a crossing with any state $p_B$ having 
$c_2^B \neq 0$ is guaranteed at sufficiently long times. This is referred 
to as the strong Mpemba effect. More generally, a 
weak Mpemba effect occurs whenever a crossing happens without the 
strict vanishing of $c_2^A$, due to a sufficiently small value of 
$|c_2^A|$ relative to $|c_2^B|$. The first controlled experimental 
demonstration of the Mpemba effect was later provided by Kumar and 
Bechhoefer in a colloidal system~\cite{Kumar2020}, confirming the theoretical 
picture of Lu and Raz.

This framework has since been extended to open quantum systems, where the Mpemba effect has been studied in the context of Lindblad~\cite{Carollo2021,Nava2019,Moroder2024} and non-Markovian dynamics~\cite{Strachan2024,Ze-Zhou2025} and has been demonstrated experimentally~\cite{Aharony2024,Joshi2024,Zhang2025}. More broadly, ``Mpemba effect'' has become an umbrella term for anomalous equilibration dynamics, encompassing phenomena such as the anomalously fast restoration of symmetries in classical~\cite{Summer2026} and quantum~\cite{Ares2023Nat,Rylands2024,Turkeshi2024,Ares2025} systems, and relaxation to nonequilibrium steady states~\cite{Nava2024,WangWang2024}. First practical applications in quantum systems have also begun to emerge, including accelerated state preparation~\cite{Westhoff2025}, rapid cooling protocols~\cite{Mondal2026}, anomalous discharging of quantum batteries~\cite{Medina2024,Li2025}, and fast qubit reset~\cite{Lejeune2026}. This slow-mode suppression is the mechanism exploited by the optimized-initialization protocol developed above.

\section*{\updated{Results}}
\subsection*{Speeding up thermodynamic computations with optimized initializations}
\label{subsec:opt:init} 

Following~\cite{Melanson2023,Whitelam2025_clock,Aifer2024}, we model the stochastic
dynamics of the displacement vector $\mathbf{x}(t)\in\mathbb{R}^d$ of the $d$
coupled oscillators constituting the thermodynamic hardware, which encode the
symmetric positive-definite matrix $\mathbf{A}$ through the quadratic potential
$V(\mathbf{x})$ (see~\cref{eq:quadratic:potential}), with an overdamped
Langevin equation of the form
\begin{equation}
    \dot{x}_i = -\mu \frac{\partial V(\mathbf{x})}{\partial x_i}
    + \sqrt{2\mu k_BT}\,\eta_i(t)\,,
    \label{eq:overdamped:langevin}
\end{equation}
where $\mu$ is the mobility and $\eta_i(t)$ are independent Gaussian white-noise processes with
$\langle \eta_i(t)\rangle=0$ and
$\langle \eta_i(t)\eta_j(t')\rangle=\delta_{ij}\delta(t-t')$.
The stochastic dynamics~\cref{eq:overdamped:langevin} is equivalently described by the Fokker-Planck~(FP) equation for the probability distribution $p(\mathbf{x},t)$~\cite{Gardiner2004}
\begin{equation}
    \partial_t p(\mathbf{x},t)
    =
    \mu \nabla \cdot \!\left[
        (\nabla V(\mathbf{x}))\, p(\mathbf{x},t)
    \right]
    +
    \mu k_BT \nabla^2 p(\mathbf{x},t)\,,
    \label{eq:fokker-planck}
\end{equation}
or, more compactly, $\partial_t p=\mathcal{L}_{\mathrm{FP}}p$, where $\mathcal{L}_{\mathrm{FP}}$ is the FP generator.
The unique stationary state is the Boltzmann distribution
$p_{\mathrm{eq}}$.

Under detailed balance, $\mathcal{L}_{\mathrm{FP}}$ is self-adjoint with
respect to the weighted inner product
\begin{equation}
\langle f,g\rangle_w
=
\int d\mathbf{x}\,
\frac{f(\mathbf{x}) g(\mathbf{x})}{p_{\mathrm{eq}}(\mathbf{x})}.
\end{equation}
Consequently, its spectrum is real and its eigenfunctions
$\Phi_{\mathbf{n}}(\mathbf{x})$ can be chosen to form an orthonormal basis with respect to this inner product
\begin{equation}
\mathcal{L}_{\mathrm{FP}}
\Phi_{\mathbf{n}}(\mathbf{x})
=
-\mu(\mathbf{n}\cdot\boldsymbol{\lambda})
\Phi_{\mathbf{n}}(\mathbf{x}),
\qquad
\langle \Phi_{\mathbf{m}}, \Phi_{\mathbf{n}} \rangle_w
=
\delta_{\mathbf{m},\mathbf{n}}.
\end{equation}
In this basis, the general solution of
\cref{eq:fokker-planck} can be written as~\cite{LiberzonBrockett2000}
\begin{equation}
p(\mathbf{x},t)
=
p_{\mathrm{eq}}(\mathbf{x})
+
\sum_{\mathbf{n}\neq \mathbf{0}}
c_{\mathbf{n}}\,
e^{-\mathbf{n}\cdot\boldsymbol{\lambda}\mu t}
\Phi_{\mathbf{n}}(\mathbf{x}),
\label{eq:fp:spectral}
\end{equation}
where $\mathbf{n}=(n_1,\dots,n_d)$ is a multi-index of nonnegative integers, $\boldsymbol{\lambda}=(\lambda_1,\dots,\lambda_d)$ are the eigenvalues of $\mathbf{A}$, ordered ascendingly $\lambda_1 \leq \lambda_2\dots \leq\lambda_d$ and $\mathbf{n}\cdot\boldsymbol{\lambda}=\sum_i n_i\lambda_i$.
The coefficients $c_{\mathbf{n}}$ are fixed by the initial distribution $p_0(\mathbf{x})$ and slow relaxation is governed by the smallest nonzero decay rate $\mu(\mathbf{n}\cdot\boldsymbol{\lambda})$.

\paragraph{Fast thermalization}
For the quadratic encodings considered here, the eigenfunctions of the FP operator can be computed exactly.
Let $\mathbf{A}=\mathbf{U}\mathbf{\Lambda}\mathbf{U}^T$ with 
$\mathbf{\Lambda}=\mathrm{diag}(\lambda_1,\dots,\lambda_d)$ and 
$\mathbf{U}=(\mathbf{u}_1, \dots, \mathbf{u}_d)$ the matrix of eigenvectors, 
and introduce normal-mode coordinates $\mathbf{y}=\mathbf{U}^T\mathbf{x}$.
In these coordinates, the FP equation decouples into a sum of independent one-dimensional Ornstein-Uhlenbeck generators
\begin{equation}
    \partial_t p(\mathbf{y},t)
    =
    \sum_{i=1}^d
    \left[
        \mu\lambda_i\partial_{y_i}(y_i p)
        +
        \mu k_BT\,\partial_{y_i}^2 p
    \right],
\end{equation}
and the equilibrium distribution factorizes as
\begin{equation}
    p_{\mathrm{eq}}(\mathbf{y})
    =
    \prod_{i=1}^d
    \sqrt{\frac{\lambda_i}{2\pi k_BT}}
    \exp\!\left(
        -\frac{\lambda_i y_i^2}{2k_BT}
    \right).
    \label{eq:in:eigenbasis}
\end{equation}
The eigenfunctions of the FP generator can be written (up to normalization) as

\begin{equation}
\Phi_{\mathbf n}(\mathbf y)
=
p_{\mathrm{eq}}(\mathbf y)
\prod_{i=1}^d
H_{n_i}\!\left(
\sqrt{\frac{\lambda_i}{k_BT}}\,y_i
\right),
\label{eq:phi:hermite}
\end{equation}
where $H_n$ denotes the Hermite polynomials.
Importantly, expectation values of quadratic observables  are determined solely by FP modes whose Hermite part has total degree
$|\mathbf{n}|=\sum_i n_i\le 2$.
Specifically, the dynamics of the covariance components
\begin{equation}
    \Sigma'_{ij}(t)\equiv\langle y_i y_j\rangle_t
\end{equation}
involve only the following quadratic FP modes:
(i) variance modes with $\mathbf{n}=(0,\dots,2,\dots,0)$, decaying with rates $2\mu\lambda_i$;
(ii) cross-covariance modes with $\mathbf{n}=(0,\dots,1,\dots,1,\dots,0)$ for $i\neq j$, decaying with rates $\mu(\lambda_i+\lambda_j)$.
The slowest quadratic FP mode is therefore the variance mode associated with $\lambda_1$, corresponding to $\mathbf{n}=(2,0,\dots,0)$ and decay rate $2\mu\lambda_1$.
The corresponding FP eigenfunction is
\begin{equation}
\begin{split}
    \Phi_{(2,0,\dots,0)}(\mathbf{y})
    &=
    p_{\mathrm{eq}}(\mathbf{y})
    H_2\left(
        \sqrt{\frac{\lambda_1}{k_BT}}\,y_1
    \right) \\
    &=
    p_{\mathrm{eq}}(\mathbf{y})
    \left(
        \frac{\lambda_1 y_1^2}{k_BT}-1
    \right).
\end{split}
\label{eq:phi:2}
\end{equation}
Projecting the initial distribution onto this mode yields, up to normalization,
\begin{align}
    c_{(2,0,\dots,0)}
    &\propto
    \langle p_0(\mathbf y),\Phi_{(2,0,\dots,0)}(\mathbf y)\rangle_w
    \nonumber\\
    &=
    \int d\mathbf y\,
    \frac{p_0(\mathbf y)\,\Phi_{(2,0,\dots,0)}(\mathbf y)}
         {p_{\mathrm{eq}}(\mathbf y)} .
    \label{eq:c2:proj:1}
\end{align}
Inserting the explicit form of the slowest quadratic FP eigenfunction from
\cref{eq:phi:2}, we obtain
\begin{align}
    c_{(2,0,\dots,0)}
    &\propto
    \int d\mathbf y\,
    p_0(\mathbf y)
    \left(\frac{\lambda_1 y_1^2}{k_BT}-1\right)
    \nonumber\\
    &=
    \frac{\lambda_1}{k_BT}
    \left(
        \langle y_1^2\rangle_0
        -
        \langle y_1^2\rangle_{\mathrm{eq}}
    \right),
    \label{eq:c2:zero}
\end{align}
where we used the normalization $\int d\mathbf{y}\,p_0(\mathbf{y})=1$ and the
equilibrium variance $\langle y_1^2\rangle_{\mathrm{eq}}=k_BT/\lambda_1$.
\updated{Thus, setting the coefficient of the slowest relaxation mode to zero requires
$\langle y_1^2\rangle_0=\langle y_1^2\rangle_{\rm eq}$, i.e., matching the initial variance along the corresponding eigenmode to its equilibrium value.}
This eliminates the dominant bottleneck in the approach to equilibrium.
For the $k$-th variance mode $n=(0,\dots,2,\dots,0)$, an analogous expression holds, with the amplitude determined by the deviation of $\langle y_k^2\rangle_0$ from its equilibrium value $k_BT/\lambda_k$.

\cref{eq:c2:zero} reflects the general Mpemba strategy of suppressing the slowest relaxation modes. Since for quadratic potentials these modes factorize along the eigenvectors of $\mathbf{A}$ according to~\cref{eq:phi:hermite}, their amplitudes are fully determined by the corresponding variances. This allows one to work with the eigenvectors of $\mathbf{A}$ rather than with the full Fokker-Planck eigenfunctions, and to formulate the optimization directly in terms of the covariance.
The dynamics of the covariance matrix
$\mathbf{\Sigma}(t)=\langle\mathbf{x}(t)\mathbf{x}(t)^T\rangle$ under~\cref{eq:overdamped:langevin} is governed by the Lyapunov equation
\begin{equation}
    \dot{\mathbf{\Sigma}}(t)=
    -\mu\bigl(\mathbf{A}\mathbf{\Sigma}(t)+\mathbf{\Sigma}(t)\mathbf{A}\bigr)
    +2\mu k_BT\,\mathbb{1},
    \label{eq:lyapunov}
\end{equation}
(see Supplementary Note 1 for a derivation) which has stationary solution $\mathbf{\Sigma}^{\mathrm{eq}}=k_BT\,\mathbf{A}^{-1}$.
Transforming to the eigenbasis of $\mathbf{A}$, $\mathbf{\Sigma}'=\mathbf{U}^T\mathbf{\Sigma}\mathbf{U}$, the dynamics decouples as
\begin{equation}
    \dot{\Sigma}'_{ij}(t)=
    -\mu(\lambda_i+\lambda_j)\Sigma'_{ij}(t)
    +2\mu k_BT\,\delta_{ij}.
\end{equation}
In particular
\begin{equation}
    \dot{\Sigma}'_{11}(t)=
    -2\mu\lambda_1
    \left(
        \Sigma'_{11}(t)-\frac{k_BT}{\lambda_1}
    \right),
\end{equation}
so that initializing $\Sigma'_{11}(0)=k_BT/\lambda_1$ makes this slowest mode stationary at all times.
Since $\Sigma'_{11} = \langle y_1^2 \rangle$, the condition 
$\Sigma'_{11}(0) = k_BT/\lambda_1$, which we refer to as the \textit{prethermalization} condition, is identical to the FP orthogonality condition~(\cref{eq:c2:zero}).

\paragraph{The hybrid digital-thermodynamic algorithm}
In terms of physical coordinates $\mathbf{x}$, prethermalizing the first $K$ quadratic FP modes corresponds to initializing the covariance matrix as
\begin{equation}
    \mathbf{\Sigma}^{\mathrm{opt}}_0(K)\equiv
    \sum_{k=1}^{K}\frac{k_BT}{\lambda_k}\,
    \mathbf{u}_k\mathbf{u}_k^T,
    \label{eq:optimized:covariance:matrix:initialization:k:modes}
\end{equation}
where $\mathbf{u}_k$ is the eigenvector of $\mathbf{A}$ associated with the \mbox{$k$-th} smallest eigenvalue.
This initialization matches the equilibrium variance along the first
$K$ eigenmodes of $\mathbf A$, while the remaining directions are
initialized with zero variance.

The partial spectral information necessary for this optimized initialization can be obtained efficiently
using Krylov-subspace methods such as the Lanczos algorithm~\cite{Lanczos1950,Saad2011}. Computing the lowest
$K$ eigenpairs has total computational cost
$\mathcal{O}\!\left(K\,\mathrm{nnz}(\mathbf A)\right)$ for sparse matrices, \updated{where $\mathrm{nnz}(A)$ denotes the number of nonzero entries of $A$,} or $\mathcal{O}(K d^2)$ for dense matrices, avoiding the
$\mathcal{O}(d^3)$ cost of full diagonalization when $K\ll d$.

While~\cref{eq:lyapunov} provides a convenient description of relaxation at the level of second moments, the experimentally controllable quantity is the initial condition $\mathbf{x}_0$ for each run.
To realize~\cref{eq:optimized:covariance:matrix:initialization:k:modes} in practice, we sample the initial conditions from a Gaussian distribution
$\mathbf{x}^\mathrm{opt}_0\sim\mathcal{N}\!\left(\mathbf{0},\mathbf{\Sigma}^{\mathrm{opt}}_0(K)\right)$,
which can be written as
\begin{equation}
    \mathbf{x}_0^{\mathrm{opt}}(K)=
    \sum_{k=1}^{K}
    \sqrt{\frac{k_BT}{\lambda_k}}\,
    z_k\,\mathbf{u}_k,
    \label{eq:optimal:sample:initial:cond}
\end{equation}
with $z_k\sim\mathcal{N}(0,1)$.
For $K=0$ this reduces to the standard choice $\mathbf{x}_0^{\mathrm{std}}=\mathbf{0}$. 
The resulting hybrid digital-thermodynamic protocol consists of two stages:
\begin{enumerate}[label=(\roman*)]
    \item On a digital processor, compute the $K$ lowest eigenpairs of $\mathbf{A}$ via the Lanczos algorithm.
    \item Sample $\mathbf{x}_0$ according to~\cref{eq:optimal:sample:initial:cond}, encode it in the thermodynamic hardware and let the system thermalize to estimate the desired matrix function $f(\mathbf{A})$ from measured observables.
\end{enumerate}
A comparison between the standard and the optimized initialization is shown pictorially in panel b) of Fig.~1.

In practice, encoding $\mathbf{x}_0^{\text{opt}}$ on analog hardware simply requires \updated{preparing the corresponding initial node-voltage vector
\(\mathbf{v}(0)=\mathbf{x}^{\rm opt}_0\) in the coupled circuits.} The energy stored in this initial 
configuration, $E_{\text{init}} =\frac{1}{2}\mathbf{v}^T \mathbf{A} \mathbf{v}$~\cite{Melanson2023}, is of order 
$k_B T$ per degree of freedom. Since the initialization depends on the eigenpairs of 
$\mathbf{A}$, which are encoded in the capacitive couplings of the hardware, deviations of 
those couplings from their nominal values will render the initialization suboptimal, making 
accurate calibration crucial.
\updated{We note that RLC-circuit implementations of thermodynamic computing can also operate in underdamped regimes, where relaxation is governed by coupled voltage- and current-like degrees of freedom. In Supplementary Note 2, we show that the optimized-initialization condition then generalizes to the full initial phase-space covariance, including voltage–voltage, current–current, and voltage–current correlations, but reduces exactly to the configurational condition used here when the current-like sector is initialized thermally and uncorrelated with the voltages.}

\paragraph{\updated{Resulting Speedups}}
During the thermalization stage, we quantify the distance from equilibrium by monitoring the deviation of the
covariance matrix
\begin{equation}
    \mathcal{E}(t)\equiv
    \|\mathbf{\Sigma}(t)-\mathbf{\Sigma}^{\mathrm{eq}}\|_F\,,
    \label{eq:cov:error}
\end{equation}
where $\|\cdot\|_F$ denotes the Frobenius norm, and define 
the time to reach thermalization threshold $\epsilon_t$,
\begin{equation}
    t_0(\epsilon_t,K)\equiv
    \inf\{t\ge 0:\ \mathcal{E}(t)\le\epsilon_t\}.
\end{equation}
For sufficiently small $\epsilon_t$, the residual error is dominated by the
slowest non-prethermalized mode, yielding
\begin{equation}
    t_0(\epsilon_t,K)\simeq
    \frac{1}{2\mu\lambda_{K+1}}
    \ln\!\left(\frac{|c_{K+1}|}{\epsilon_t}\right),
    \label{eq:crossing:time}
\end{equation}
where $c_{K+1}$ is the amplitude of the first non-prethermalized quadratic covariance mode. Then, we can define an associated Mpemba speedup factor as
\begin{equation}
    S(\epsilon_t,K)\equiv
    \frac{t_0(\epsilon_t,0)}{t_0(\epsilon_t,K)}
    \xrightarrow{\epsilon_t\to 0}
    \frac{\lambda_{K+1}}{\lambda_1}\,.
    \label{eq:mpemba:speedup}
\end{equation}

\updated{It is convenient to characterize the digital preprocessing by the fraction $\alpha=K/d$ of spectral modes treated digitally. The useful regime of the
protocol is $\alpha\ll 1$: the speedup grows with $K$, but so does the
digital preprocessing cost. In the opposite limit $\alpha\to 1$, the
preprocessing approaches a full spectral calculation, at which point the analog
stage becomes redundant. The resulting trade-off is
analyzed in Supplementary Note 3.}

We also emphasize that the optimized initializations introduced here accelerate the thermalization stage of thermodynamic computations. For single-stage algorithms such as matrix inversion and the solution of linear systems, which (in their ensemble-averaging variants) terminate once equilibrium is reached, this results in a speedup of the full algorithm. By contrast, in multi-stage algorithms including determinant estimation or matrix exponentiation, the computation separates into a thermalization process followed by an additional post-thermalization stage, which is not affected by the optimized initialization.

\subsection*{\updated{Applications}}
\label{sec:examples}
In the following, we illustrate the effectiveness of the hybrid digital-thermodynamic protocol with Mpemba-inspired optimized initializations for two important matrix operations. We simulate the thermalization dynamics~\cref{eq:overdamped:langevin,eq:lyapunov} numerically and compare the obtained speedups to the theoretical predictions.
To highlight the role of the spectrum of $\mathbf{A}$, we consider two ensembles that represent opposite regimes: one 
where the spectral range grows linearly with $d$, and one where it remains 
$O(1)$.
Specifically, for the first case, we construct matrices with a fixed spectrum
\begin{equation}
\mathrm{spec}
=
\bigl(
\lambda_{\min},
\lambda_{\min}+\delta,
\dots,
\lambda_{\min}+(d-1)\delta
\bigr),
\label{eq:spec:linear:spacing}
\end{equation}
and Haar-random eigenvectors,
\begin{equation}
\mathbf{A}_{\mathrm{f}}
=
\mathbf{U}\,
\mathrm{diag}(\mathrm{spec})\,
\mathbf{U}^T,
\qquad
\mathbf{U}\sim \mathrm{Haar}(O(d)).
\label{eq:haar:random:eigenv}
\end{equation}
By construction, the level spacing for this ensemble is fixed to $\delta$ and independent of the matrix dimension~$d$.
As a second ensemble, we consider positive Wishart matrices, which have been routinely employed in thermodynamic computing benchmarks~\cite{Aifer2024,Melanson2023,Duffield2025}, defined as
\begin{equation}
\mathbf{A}_\mathrm{w}
=
\frac{1}{m}\,\mathbf{X}^T\mathbf{X},
\quad
\mathbf{X}\in\mathbb{R}^{m\times d},
\quad
X_{ij}\sim\mathcal{N}(0,1)\,,
\label{eq:wishart:def}
\end{equation}
with $m\ge d$.
For fixed aspect ratio $c=m/d>1$, the eigenvalues follow the Marchenko-Pastur law~\cite{Marcenko_1967, BaiSilverstein2010} with support \begin{equation}
\lambda_{\min} = \left(1 - c^{-1/2}\right)^2, 
\quad
\lambda_{\max} = \left(1 + c^{-1/2}\right)^2 .
\label{eq:mp_support}
\end{equation}
Thus, the spectral interval is $O(1)$ and independent of~$d$, implying a typical mean level spacing $\sim 1/d$ and a spectral density that grows proportionally to $d$.

\updated{The Wishart and linear-spectrum Haar ensembles above follow the
conventions of the thermodynamic-computing literature and have
relatively flat spectra. The Bayesian machine-learning applications
that partially motivate this work, however, are dominated by regularized kernel
Gram matrices~\cite{RasmussenWilliams2006} whose spectra
decay rapidly (near-exponentially for smooth kernels and polynomially
for rougher ones). This is precisely the regime in which the asymptotic
speedup $\lambda_{K+1}/\lambda_1$ grows fastest with $K$, so the
protocol is expected to be most effective there. To probe this regime, we consider a third benchmark: the
radial-basis-function (RBF) kernel Gram ensemble, discussed in Supplementary Note 4.}

\updated{We emphasize that the digital stage runs once per matrix, whereas the
thermalization is repeated over the $N_\mathrm{traj}$ trajectories used to reach the target accuracy. The relative cost of the two stages is therefore governed by $N_\mathrm{traj}$ and by the hardware time constant $\tau_c$ that sets the physical duration of the analog relaxation. We analyze this trade-off in Supplementary Note 3, where the optimal number of prethermalized modes $K^\star(\tau_c)$ follows from balancing the digital preprocessing against the analog thermalization time.}
\paragraph{Faster matrix inversion}
%
First, we study thermodynamic matrix inversion~\cite{Aifer2024}. We initialize the system according to~\cref{eq:optimized:covariance:matrix:initialization:k:modes} and solve~\cref{eq:lyapunov} numerically.
\begin{figure}[t] 
    \centering
    \subfloat[\label{subfig:matrix:inv:a}]{
        \includegraphics[width=0.95\columnwidth]{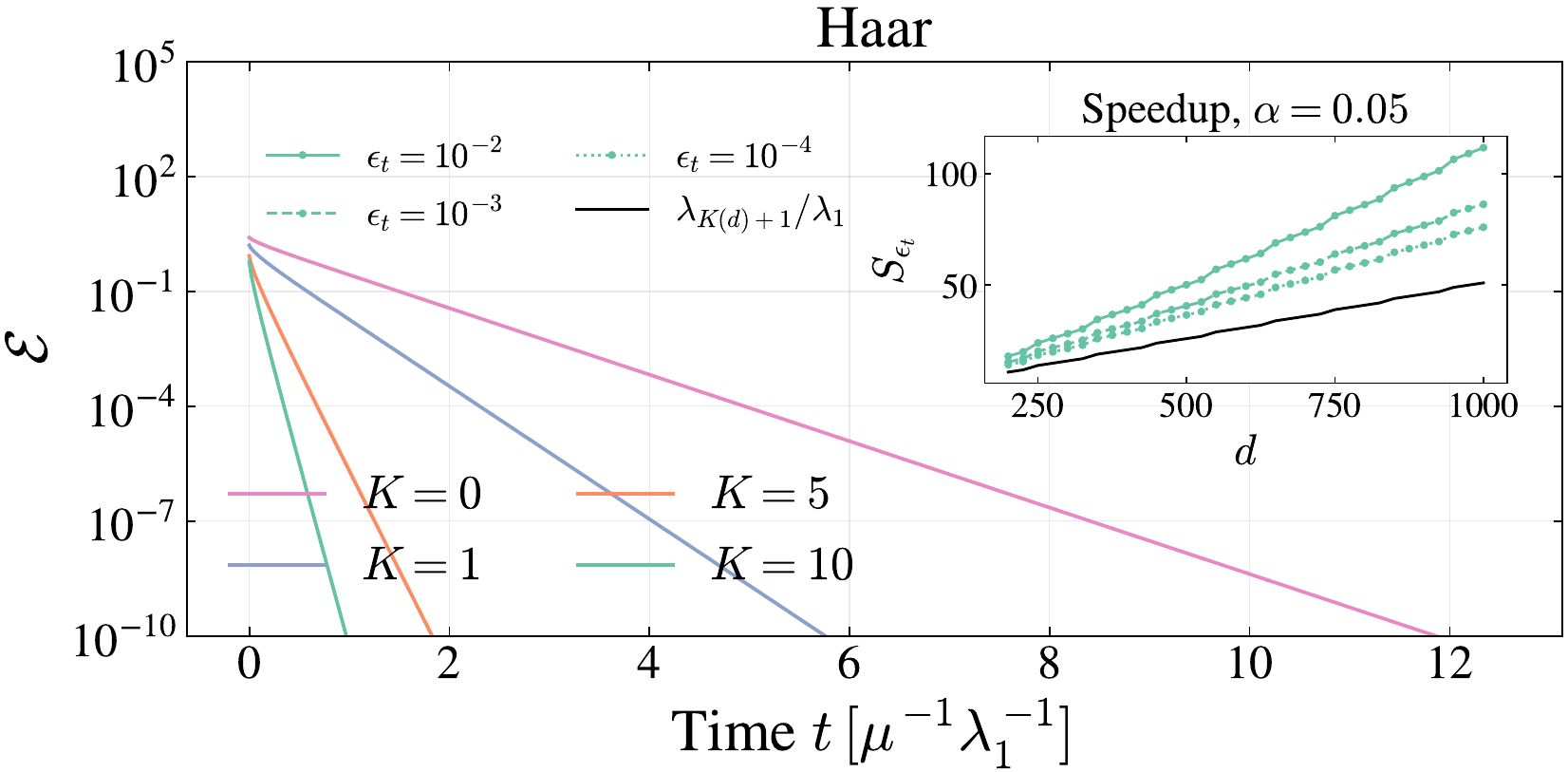}
    }
    \vspace{-0.4cm} 
    \subfloat[\label{subfig:matrix:inv:b}]{
        \includegraphics[width=0.95\columnwidth]{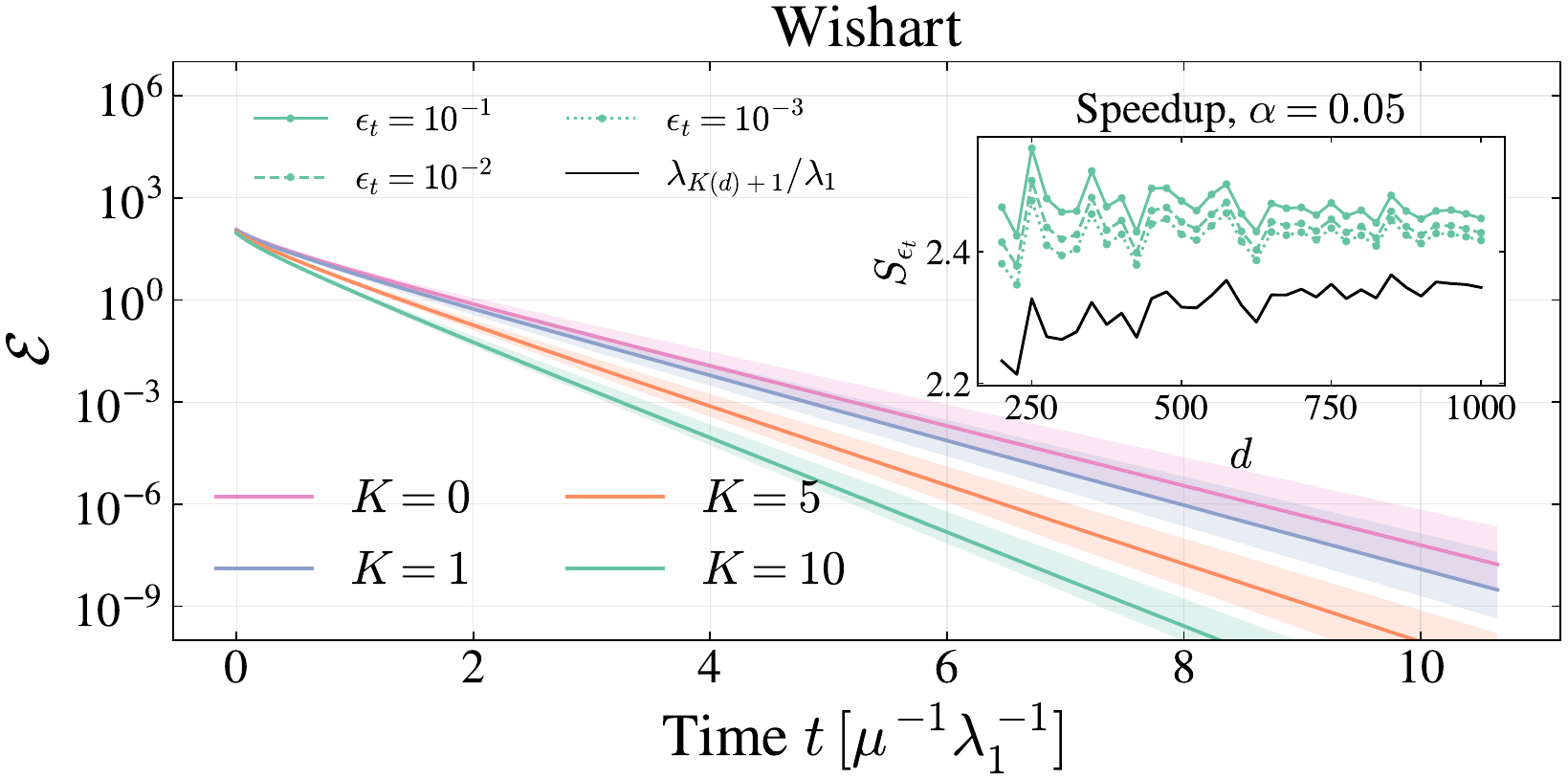}
    \hspace{-0.1cm}
    }
    \caption{\textbf{Accelerating thermodynamic matrix inversion}.
      (a) Ensemble with linearly spaced eigenvalues and Haar-random eigenvectors~\cref{eq:spec:linear:spacing,eq:haar:random:eigenv}.
    (b) Positive Wishart random matrices~\cref{eq:wishart:def}.
    Both panels show the time-dependent Frobenius distance between the evolving covariance matrix and the exact inverse~\cref{eq:cov:error} for $d=500$.
    The parameter $K$ denotes the number of slow modes that are prethermalized in the optimized initialization~\cref{eq:optimized:covariance:matrix:initialization:k:modes}.
    Insets show the resulting speedups for \updated{$\alpha=K/d=0.05$} as a function of the matrix dimension $d$, evaluated at three different thermalization error thresholds $\epsilon_t$.
    Black lines indicate the corresponding asymptotic speedup~\cref{eq:mpemba:speedup}.
    We set $\delta=0.5$, \updated{$m=1.5d$}, $\mu=k_BT=1$, and average over \updated{$100$} random matrices for each panel. The shaded regions indicate the full min-max spread over the sampled ensemble.}
\label{fig:matrix:inversion}
\end{figure}
Fig.~2 illustrates the effect of Mpemba-optimized initializations on the relaxation of the covariance matrix. Increasing the number $K$ of prethermalized slow modes leads to a clear acceleration of the dynamics in both ensembles.
The different long-time decay behaviors visible in the main panels reflect the progressive suppression of slow relaxation modes.
The insets quantify the corresponding speedups as a function of the matrix dimension.
For fixed \updated{$\alpha$}, upon decreasing the target error threshold $\epsilon_t$ the speedups converge toward the asymptotic spectral prediction~\cref{eq:mpemba:speedup}.
This behavior reflects the fact that, at sufficiently small errors, the relaxation is ultimately controlled by a single remaining slow mode.
A qualitative difference emerges between the two ensembles.
For the fixed-spectrum case (panel (a)), the speedups \updated{grow linearly with} the matrix dimension, reflecting the fact that the eigenvalue spacing does not depend on $d$.
In contrast, for Wishart matrices, panel~(b) shows that the speedup
\updated{remains approximately independent of $d$ at fixed
$\alpha=K/d$. Indeed, as $d$ increases the eigenvalues become more
densely packed within the Marchenko-Pastur support
\cref{eq:mp_support}; however, because $K=\alpha d$ also grows linearly
with $d$, the cutoff eigenvalue $\lambda_{K+1}$ approaches a fixed
spectral quantile. As a result, the ratio controlling the speedup remains
approximately constant with $d$.}

We also note that adding a linear term $-\mathbf b^{T}\mathbf x$ to the quadratic potential~\cref{eq:quadratic:potential} shifts the equilibrium mean to $\langle \mathbf x\rangle_{\mathrm{eq}}=\mathbf A^{-1}\mathbf b$, enabling the solution of linear systems via sampling of the equilibrium mean~\cite{Aifer2024}. Since the relaxation dynamics are controlled by the same spectrum of $\mathbf A$, the Mpemba speedups discussed here apply equally to linear system solving and to matrix inversion.
\updated{The protocol extends directly to ridge-regularized
matrix inversion~\cite{Hoerl01021970}. In applications where $\mathbf{A}$ is ill-conditioned, one may
encode
\begin{equation}
\mathbf{A}_{\rm reg}=\mathbf{A}+\epsilon_{\rm reg}\mathbb{1},
\qquad \epsilon_{\rm reg}>0,
\end{equation}
instead of $\mathbf{A}$, thereby estimating the regularized inverse
$(\mathbf{A}+\epsilon_{\rm reg}\mathbb{1})^{-1}$. Since this transformation leaves
the eigenvectors unchanged and shifts the eigenvalues as
\(\lambda_i\mapsto \lambda_i+\epsilon_{\rm reg}\), the optimized
initialization becomes
\begin{equation}
\mathbf{\Sigma}^{\rm opt}_{0,{\rm reg}}(K)
=
\sum_{k=1}^{K}
\frac{k_B T}{\lambda_k+\epsilon_{\rm reg}}
\mathbf{u}_k \mathbf{u}_k^T .
\end{equation}
The corresponding asymptotic speedup is therefore
\begin{equation}
S_{\rm reg}(K)
=
\frac{\lambda_{K+1}+\epsilon_{\rm reg}}
{\lambda_1+\epsilon_{\rm reg}} \geq 1.
\end{equation}
Thus, ridge regularization increases the slowest covariance relaxation
rate from \(2\mu\lambda_1\) to
\(2\mu(\lambda_1+\epsilon_{\rm reg})\), improving conditioning and
mitigating near-zero eigenvalues. At the same time, because the same positive shift is applied to all eigenvalues, ridge regularization reduces the relative separation of the low-lying relaxation rates. It therefore trades some relative Mpemba speedup for increased robustness and faster absolute relaxation, according to
\begin{equation}
\frac{d S_{\rm reg}(K)}{d\epsilon_{\rm reg}}
=
\frac{\lambda_1-\lambda_{K+1}}
{(\lambda_1+\epsilon_{\rm reg})^2}
\leq 0.
\end{equation}
}

Finally, we mention that rescaling $\mathbf{A} \to \chi \mathbf{A}$ by $\chi > 1$ speeds up thermalization by the same factor, since the relaxation rates scale as $\mu\lambda_i \to \chi\mu\lambda_i$, while the output is recovered by rescaling the result by $\chi^{-1}$ (or equivalently by rescaling $\mathbf{b} \to \chi\mathbf{b}$ for 
linear systems, leaving the solution unchanged). Within the experimental constraints 
on achievable coupling strengths (set in current 
hardware by the discrete and bounded set of available capacitor values~\cite{Melanson2023}), this might provide a simple and 
complementary route to accelerating thermodynamic computations. 
A related strategy, based on rescaling the potential while simultaneously injecting additional noise to preserve the equilibrium distribution, has been explored in 
Ref.~\cite{Whitelam2025_clock}.

\paragraph{Faster matrix-determinant computation}
%
Another fundamental linear algebra operation that can be implemented on thermodynamic hardware is the computation of matrix determinants
\cite{Aifer2024}. 
The key idea is to express the determinant as an equilibrium free energy difference, which can then be estimated from nonequilibrium work statistics using fluctuation relations.
Consider two quadratic potentials $V_{\mathbf{A}}(\mathbf{x})$ and
$V_{\mathbf{B}}(\mathbf{x})$ of the form given in
\cref{eq:quadratic:potential}, encoded by two symmetric positive definite
matrices $\mathbf{A}$ and $\mathbf{B}$. The equilibrium free energy associated
with $V_{\mathbf{A}}$ at temperature $T$ is
$F_{\mathbf{A}} = -k_B T \log Z_{\mathbf{A}}$, where
$Z_{\mathbf{A}} = \int d\mathbf{x}\, \exp(-V_{\mathbf{A}}(\mathbf{x})/k_B T)$.
Since $Z_{\mathbf{A}} \propto (\det \mathbf{A})^{-1/2}$, the free-energy
difference
\begin{equation}
\Delta F \equiv F_{\mathbf{B}} - F_{\mathbf{A}}
= \frac{k_B T}{2}
\log\!\left(\frac{\det \mathbf{B}}{\det \mathbf{A}}\right)
\end{equation}
is directly related to the ratio of determinants. Rearranging yields
\begin{equation}
\det(\mathbf{A}) =
e^{-\frac{2}{k_B T}\Delta F}\,\det(\mathbf{B}) .
\label{eq:det-from-free-energy}
\end{equation}
Therefore, if $\det(\mathbf{B})$ is known, computing $\det(\mathbf{A})$ reduces to estimating the equilibrium free-energy difference $\Delta F$.

Here, we estimate $\Delta F$ using Crooks' fluctuation theorem~\cite{Crooks1999}. While Jarzynski's equality~\cite{Jarzynski1997} yields an unbiased estimator 
of $\Delta F$, it converges slowly in practice because it is 
dominated by rare trajectories. 
We consider a nonequilibrium protocol that transforms the system from potential $V_{\mathbf{A}}$ to $V_{\mathbf{B}}$ over a finite time $\tau$. Let $W$ denote the work performed during such a forward process, and let $W'$ denote the work measured during the corresponding reverse protocol transforming $V_{\mathbf{B}}$ back to $V_{\mathbf{A}}$. Crooks' theorem states that the probability distributions of forward and reverse work satisfy
\begin{equation}
\frac{P_{\mathrm{F}}(W)}{P_{\mathrm{R}}(-W)}
=
\exp\!\left(\frac{W-\Delta F}{k_B T}\right).
\end{equation}

In practice, the free energy difference $\Delta F$ is obtained from the
sampled forward and reverse work values using the Bennett acceptance ratio
(BAR) estimator~\cite{Bennett1976}.  For equal numbers of forward and reverse
trajectories $N_\mathrm{traj}$, $\Delta F$ is defined as the solution of the nonlinear equation
\begin{equation}
\begin{split}
&\sum_{i=1}^{N_\mathrm{traj}}
\frac{1}{1+\exp[(W_i-\Delta F)/k_B T]}
=
 \\ &\sum_{j=1}^{N_\mathrm{traj}}
\quad
\frac{1}{1+\exp[(\Delta F-W'_j)/k_B T]} .
\end{split}
\label{eq:bar}
\end{equation}
Once $\Delta F$ has been obtained, the
determinant follows directly from \cref{eq:det-from-free-energy}.
The nonequilibrium protocol is implemented by introducing a time-dependent interpolation
$\mathbf{A}(t) = (1-s(t))\,\mathbf{A} + s(t)\,\mathbf{B}$ with $s(0)=0$ and
$s(\tau)=1$, which we take to increase linearly in time. Along a single Langevin trajectory $\mathbf{x}(t)$, the work is
given by
\begin{equation}
W =
\int_0^\tau dt\;
\frac{1}{2}\,
\mathbf{x}(t)^T \dot{\mathbf{A}}(t)\,\mathbf{x}(t) .
\label{eq:work}
\end{equation}
Thus, thermodynamic determinant estimation proceeds by first allowing
the system to thermalize under $V_{\mathbf{A}}(\mathbf{x})$ ($V_{\mathbf{B}}(\mathbf{x})$) and then applying
the forward (reverse) protocol $\mathbf{A}\!\to\!\mathbf{B}$ ($\mathbf{B}\!\to\!\mathbf{A}$) to generate work realizations
and estimate the free-energy difference using \cref{eq:bar}. 
Note that Mpemba-type optimized initializations affect only the thermalization stage by reducing the time required to reach equilibrium, while the subsequent
work-sampling stage and its statistical convergence remain unchanged.

\begin{figure}[t] 
    \centering
    \subfloat[\label{subfig:matrix:det:a}]{
        \includegraphics[width=0.95\columnwidth]{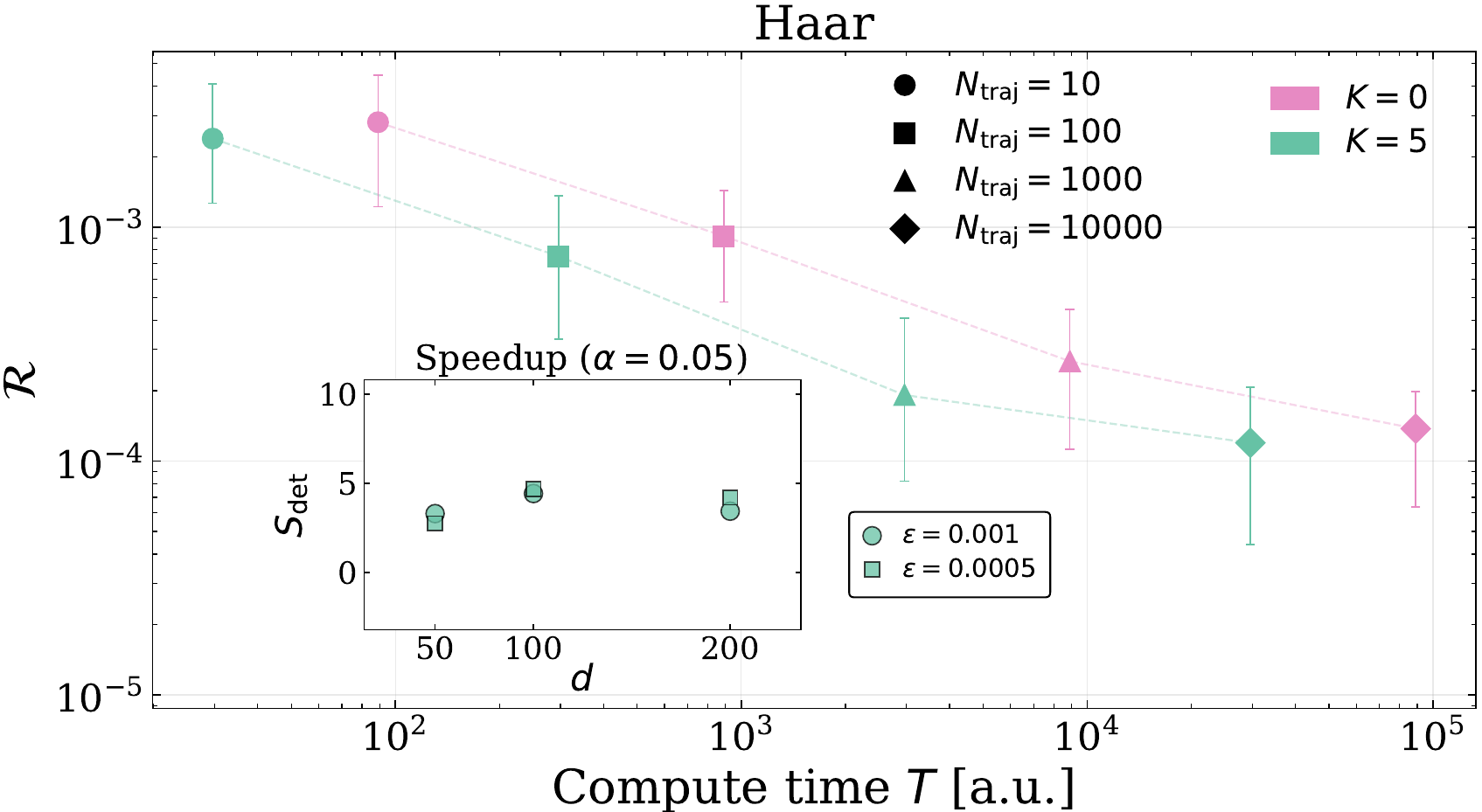}
    }
    \vspace{-0.4cm} 
    \subfloat[\label{subfig:matrix:det:b}]{
        \includegraphics[width=0.95\columnwidth]{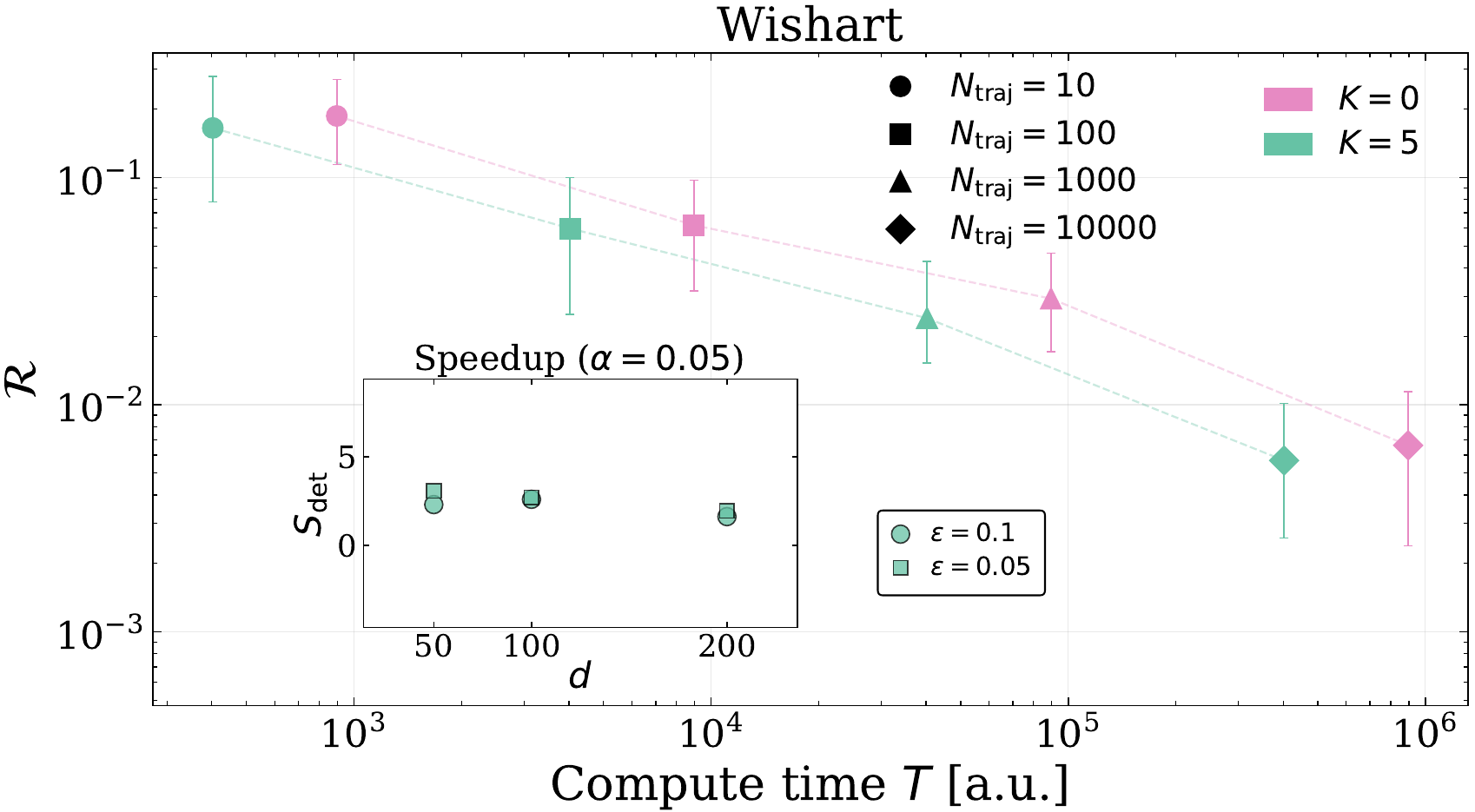}
    \hspace{-0.1cm}
    }
\label{fig:matrix:determinant}
\caption{\textbf{Accelerating thermodynamic matrix-determinant computation}. (a) Matrices with a linearly spaced spectrum and Haar-random eigenvectors,~\cref{eq:spec:linear:spacing,eq:haar:random:eigenv}. (b) Random positive Wishart matrices, \cref{eq:wishart:def}. The main panels show the relative log-determinant error~\cref{eq:rel:log:error} as a function of the total compute time $T$~\cref{eq:total:compute:time}, comparing standard initializations ($K=0$) with Mpemba-optimized initializations (\updated{$K=5$}) for different numbers of trajectories. Insets show the resulting speedup factor $S_\mathrm{det}$ as a function of the matrix dimension $d$ \updated{ for $\alpha=K/d=0.05$}, where the total time required to reach the target error $\epsilon_d$ is obtained by log-log interpolation of the median error curve between the two bracketing $N_{\mathrm{traj}}$ values. \updated{For the thermalization time $t_0$ in \cref{eq:thermalization:time:det},
we set $\epsilon_t=10^{-3}$. For the $d=100$ data shown in the main
panels, the average values are $t_0(K=0)\simeq 6.91$ and
$t_0(K=5)\simeq 0.96$ for panel~(a), and
$t_0(K=0)\simeq 88.64$ and $t_0(K=5)\simeq 38.72$ for panel~(b).} We use $\tau =2$, $\delta = 0.5$, $m = 1.5d$, $\mu = k_BT = 1$, and \updated{$dt = 10^{-2}$}. Results are averaged over \updated{$100$} independent random matrix realizations per panel; error bars indicate the spread across realizations and markers denote the median.}
\end{figure}
\updated{We now quantify the effect of the optimized initializations for this algorithm.
Before introducing the determinant-estimation error, we first estimate the
duration of the equilibration stage. For the states defined by~\cref{eq:optimal:sample:initial:cond}, the
initial covariance residual is the normalized Frobenius-norm error between
the optimized initial covariance $\mathbf{\Sigma}^{\rm opt}_0(K)$ and the equilibrium
covariance $\mathbf{\Sigma}^{\rm eq}$,
\begin{equation}
E_0(K)
\equiv
\frac{\left\|\mathbf{\Sigma}^{\rm opt}_0(K)-\mathbf{\Sigma}^{\rm eq}\right\|_F}
{\left\|\mathbf{\Sigma}^{\rm eq}\right\|_F}
=
\left(
\frac{\sum_{i=K+1}^d \lambda_i^{-2}}
{\sum_{i=1}^d \lambda_i^{-2}}
\right)^{1/2}.
\label{eq:covariance:residual}
\end{equation}}

We consider Langevin trajectories obeying~\cref{eq:overdamped:langevin}, which are evolved until \updated{the normalized covariance residual} falls below a prescribed threshold $\epsilon_t$. 
Assuming that the dynamics are dominated by the slowest remaining mode, this yields the estimate
\begin{equation}
t_0(\epsilon_t,K)
\simeq
\frac{1}{2\mu\lambda_{K+1}}
\log\!\left(\max\left(\frac{E_0(K)}{\epsilon_t},1\right)\right)
\label{eq:thermalization:time:det}
\end{equation}
for the thermalization time, after which the nonequilibrium driving protocol~\cref{eq:work} is applied.

Following~\cite{Aifer2024}, we quantify the accuracy of the determinant estimate by introducing $L \equiv \log\det(\mathbf{A})$ 
and defining the relative log-determinant error
\begin{equation}
    \mathcal{R} \equiv 
    \frac{|L - \hat{L}|}{|L|},
    \label{eq:rel:log:error}
\end{equation}
where $\hat{L}$ is the BAR estimate of $L$ (see~\cref{eq:det-from-free-energy,eq:bar}).
The total compute time to 
reach a target determinant accuracy $\epsilon_\mathrm{d}$ is
\begin{equation}
    T(\epsilon_\mathrm{d}, \epsilon_t, K) = 
    N_\mathrm{traj}(\epsilon_\mathrm{d},\epsilon_t, \tau) 
    \cdot\left[ t_0(\epsilon_t, K) + \tau \right ],
    \label{eq:total:compute:time}
\end{equation}
with $N_\mathrm{traj}(\epsilon_\mathrm{d},\epsilon_t, \tau)$ the number 
of trajectories required to achieve the accuracy $\epsilon_d$ 
for a given switching time and thermalization error. 
The corresponding speedup factor
\begin{align}
S_\mathrm{det}(\epsilon_t, K) 
&= \frac{T(\epsilon_\mathrm{d}, \epsilon_t, 0)}
        {T(\epsilon_\mathrm{d}, \epsilon_t, K)}
\notag\\
&\approx \frac{t_0(\epsilon_t, 0) + \tau}
       {t_0(\epsilon_t, K) + \tau}
\!\begin{tikzpicture}[baseline=1.31cm]
  \draw[-{Stealth}] (-0.1, 1.4) -- (0.9, 1.85) 
      node[right] {\small $S(\epsilon_t,K)$};
  \node[rotate=24.2] at (0.32, 1.82) 
      {\scriptsize $\tau \ll t_0$};
  \draw[-{Stealth}] (-0.1, 1.4) -- (0.9, 0.95) 
      node[right] {\small $1$};
  \node[rotate=-24.2] at (0.32, 1.04) 
      {\scriptsize $\tau \gg t_0$};
\end{tikzpicture}
\label{eq:speedup:determinant}
\end{align}
is approximately independent of $\epsilon_\mathrm{d}$ and $N_\mathrm{traj}$ 
for fixed $\tau$, provided $\epsilon_t$ is sufficiently small, and interpolates between the spectral Mpemba 
speedup~\cref{eq:mpemba:speedup} when thermalization dominates 
($\tau \ll t_0$) and no speedup when the switching protocol 
dominates ($\tau \gg t_0$).

Figure~3 shows the numerical simulation of thermodynamic matrix-determinant computation. 
We set $\mathbf{B} = \mathrm{tr}(\mathbf{A})/d \,\mathbb{1}$ and employ the Mpemba-type optimization~\cref{eq:optimized:covariance:matrix:initialization:k:modes} only to the forward process.
In both 
ensembles, the optimized initialization (\updated{$K = 5$}) leads to a 
clear reduction in the relative error at fixed total compute 
time, compared to the 
standard initialization ($K = 0$). 
\updated{The insets show that, at fixed
$\alpha=K/d$, the determinant speedup is approximately independent of $d$ for
both ensembles. For Wishart matrices this follows from the fact that increasing
$d$ makes the eigenvalues denser within the Marchenko-Pastur
support~\cref{eq:mp_support}, while $K=\alpha d$ selects an approximately fixed
spectral quantile. Thus the relevant cutoff eigenvalue $\lambda_{K+1}$ remains
approximately fixed. For the fixed-spectrum Haar ensemble, $\lambda_{K+1}$
increases with $d$, but the determinant protocol measures the speedup in total
compute time, including the finite driving time and the number of trajectories
needed to reach the target accuracy. These contributions largely saturate the
gain over the dimensions considered here. The residual dependence on
$\epsilon_d$ and the small fluctuations with $d$ are due to finite-sampling
effects, both in the number of trajectories and in the averaging over random
matrices.}

We note that recent works~\cite{Whitelam2025_clock, Basak2026} have shown that the accuracy of the free-energy estimate can be further improved by rescaling the potential while simultaneously 
injecting additional noise into the system. 
This strategy is complementary to the optimized initializations proposed here and could in principle be combined to achieve further gains.

\section*{\updated{Discussion}}
\label{sec:conclusion}
We introduced a hybrid approach for accelerating matrix operations, in which a digital processor efficiently computes Mpemba-optimized initializations that are encoded into thermodynamic hardware to suppress slow relaxation modes and reduce computation time. 
We derived analytical predictions for the achievable speedups and validated them with numerical simulations for two important single-stage and multi-stage thermodynamic algorithms, namely matrix inversion and determinant computation.
Overall, we observe substantial speedups across the considered tasks. The effectiveness of optimized initialization is governed by the spectral structure of the encoded matrix: larger speedups arise when the low-lying eigenvalues are well separated, whereas increasingly dense spectra gradually reduce the achievable acceleration.
More broadly, these results demonstrate that anomalous thermalization can be engineered as a computational 
resource, opening a new interface between nonequilibrium 
thermodynamics and algorithm design.

As a notable exception, we remark that the thermodynamic matrix exponentiation algorithm of Ref.~\cite{Duffield2025}, which is based on sampling two-time correlation functions, constitutes a special case of a multi-stage algorithm where 
Mpemba-optimized initializations are not beneficial. 
\updated{Although the Fokker-Planck decay rates are governed by the matrix eigenvalues, this algorithm uses a different thermodynamic encoding from the matrix-inversion protocol considered above. In particular, the stationary covariance of the relevant Ornstein-Uhlenbeck process is isotropic, $\Sigma^{\rm eq}\propto \mathbb{1}$, while the matrix $\mathbf{A}$ is encoded in the time-lagged correlations from which the matrix exponential is extracted. The stationary distribution can therefore be sampled trivially, either by drawing $d$ independent Gaussians on a digital processor, or by thermalizing $d$ identical uncoupled oscillators on thermodynamic hardware.}

Several open questions remain. Owing to its simplicity, the 
protocol can be readily implemented in existing experimental 
thermodynamic computing hardware~\cite{Melanson2023}. In this 
context, it will be essential to investigate its performance 
under realistic experimental conditions, including imperfect 
initialization and noise in the encoded couplings.
Another important direction is whether analogous initialization-based 
speedups can be achieved beyond quadratic encodings, where the 
dynamics is no longer governed by a closed Lyapunov equation 
and genuinely nonlinear relaxation effects 
arise~\cite{Freitas2021}. Finally, it is natural to ask how 
these ideas extend to quantum thermodynamic computing 
platforms~\cite{Lipka-Bartosik2024}, where the interplay 
between coherent and dissipative dynamics may qualitatively 
modify relaxation behavior and initialization strategies.

\section*{Data availability}
The data that support the findings of this study are available 
\href{https://github.com/MattiaMoroder/data_and_code_for_publication_digitally_optimized_initializations_for_fast_thermodynamic_computing.git}{here}.

\section*{\updated{Code availability}}
The code used to generate the results and reproduce the figures is available \href{https://github.com/MattiaMoroder/data_and_code_for_publication_digitally_optimized_initializations_for_fast_thermodynamic_computing.git}{here}.

\section*{Acknowledgements}
We thank Philipp Westhoff for useful comments on a previous version of the manuscript.
MM acknowledges funding from the Royal Society, Research Ireland and the European Union’s Horizon research and innovation programme under the Marie Skłodowska-Curie grant agreement No 101264565.
JG acknowledges funding from the Royal Society and Research Ireland. F.C.B. acknowledges support from Taighde Éireann - Research Ireland under grant number IRCLA/2022/3922. 
%

%


\bibliography{Literature}

\begin{thebibliography}{10}
\expandafter\ifx\csname url\endcsname\relax
  \def\url#1{\burl{#1}}\fi
\expandafter\ifx\csname urlprefix\endcsname\relax\def\urlprefix{URL }\fi
\providecommand{\bibinfo}[2]{#2}
\providecommand{\eprint}[2][]{\url{#2}}
\providecommand{\doi}[1]{\url{https://doi.org/#1}}
\bibcommenthead

\bibitem{Conte2019}
\bibinfo{author}{Conte, T.} \emph{et~al.}
\newblock \bibinfo{title}{Thermodynamic computing} (\bibinfo{year}{2019}).
\newblock \urlprefix\url{https://arxiv.org/abs/1911.01968}.
\newblock
  \bibinfo{eprint}{{\href{https://arxiv.org/abs/1911.01968}{{arXiv:1911.01968}}}}.

\bibitem{Wolpert2024}
\bibinfo{author}{Wolpert, D.~H.} \emph{et~al.}
\newblock \bibinfo{title}{Is stochastic thermodynamics the key to understanding
  the energy costs of computation?}
\newblock \emph{\bibinfo{journal}{Proceedings of the National Academy of
  Sciences}} \textbf{\bibinfo{volume}{121}}, \bibinfo{pages}{e2321112121}
  (\bibinfo{year}{2024}).
\newblock \urlprefix\url{https://www.pnas.org/doi/abs/10.1073/pnas.2321112121}.

\bibitem{Aifer2024}
\bibinfo{author}{Aifer, M.} \emph{et~al.}
\newblock \bibinfo{title}{Thermodynamic linear algebra}.
\newblock \emph{\bibinfo{journal}{npj Unconventional Computing}}
  \textbf{\bibinfo{volume}{1}}, \bibinfo{pages}{13} (\bibinfo{year}{2024}).
\newblock
  \urlprefix\url{https://www.nature.com/articles/s44335-024-00014-0#citeas}.

\bibitem{Melanson2023}
\bibinfo{author}{Melanson, D.} \emph{et~al.}
\newblock \bibinfo{title}{Thermodynamic computing system for {AI}
  applications}.
\newblock \emph{\bibinfo{journal}{Nature Communications}}
  \textbf{\bibinfo{volume}{16}}, \bibinfo{pages}{3757} (\bibinfo{year}{2025}).
\newblock
  \urlprefix\url{https://www.nature.com/articles/s41467-025-59011-x#citeas}.

\bibitem{jelincic2025}
\bibinfo{author}{Jelinčič, A.} \emph{et~al.}
\newblock \bibinfo{title}{An efficient probabilistic hardware architecture for
  diffusion-like models} (\bibinfo{year}{2025}).
\newblock \urlprefix\url{https://arxiv.org/abs/2510.23972}.
\newblock
  \bibinfo{eprint}{{\href{https://arxiv.org/abs/2510.23972}{{arXiv:2510.23972}}}}.

\bibitem{Rolandi2026}
\bibinfo{author}{Rolandi, A.} \emph{et~al.}
\newblock \bibinfo{title}{Energy-time-accuracy tradeoffs in thermodynamic
  computing} (\bibinfo{year}{2026}).
\newblock \urlprefix\url{https://arxiv.org/abs/2601.04358}.
\newblock
  \bibinfo{eprint}{{\href{https://arxiv.org/abs/2601.04358}{{arXiv:2601.04358}}}}.

\bibitem{Bartosik2024}
\bibinfo{author}{Bartosik, P.-L.} \emph{et~al.}
\newblock \bibinfo{title}{Thermodynamic algorithms for quadratic programming}
  (\bibinfo{year}{2024}).
\newblock \urlprefix\url{https://arxiv.org/abs/2411.14224}.
\newblock
  \bibinfo{eprint}{{\href{https://arxiv.org/abs/2411.14224}{{arXiv:2411.14224}}}}.

\bibitem{Duffield2025}
\bibinfo{author}{Duffield, S.}, \bibinfo{author}{Aifer, M.},
  \bibinfo{author}{Crooks, G.}, \bibinfo{author}{Ahle, T.} \&
  \bibinfo{author}{Coles, P.~J.}
\newblock \bibinfo{title}{Thermodynamic matrix exponentials and thermodynamic
  parallelism}.
\newblock \emph{\bibinfo{journal}{Phys. Rev. Res.}}
  \textbf{\bibinfo{volume}{7}}, \bibinfo{pages}{013147} (\bibinfo{year}{2025}).
\newblock
  \urlprefix\url{https://link.aps.org/doi/10.1103/PhysRevResearch.7.013147}.

\bibitem{Whitelam2026}
\bibinfo{author}{Whitelam, S.}
\newblock \bibinfo{title}{Generative thermodynamic computing}.
\newblock \emph{\bibinfo{journal}{Phys. Rev. Lett.}}
  \textbf{\bibinfo{volume}{136}}, \bibinfo{pages}{037101}
  (\bibinfo{year}{2026}).
\newblock \urlprefix\url{https://link.aps.org/doi/10.1103/kwyy-1xln}.

\bibitem{Lu2017}
\bibinfo{author}{Lu, Z.} \& \bibinfo{author}{Raz, O.}
\newblock \bibinfo{title}{Nonequilibrium thermodynamics of the markovian mpemba
  effect and its inverse}.
\newblock \emph{\bibinfo{journal}{Proc. Natl. Acad. Sci. U. S. A.}}
  \textbf{\bibinfo{volume}{114}}, \bibinfo{pages}{5083--5088}
  (\bibinfo{year}{2017}).
\newblock \urlprefix\url{https://www.pnas.org/doi/abs/10.1073/pnas.1701264114}.

\bibitem{Klich2019}
\bibinfo{author}{Klich, I.}, \bibinfo{author}{Raz, O.},
  \bibinfo{author}{Hirschberg, O.} \& \bibinfo{author}{Vucelja, M.}
\newblock \bibinfo{title}{Mpemba index and anomalous relaxation}.
\newblock \emph{\bibinfo{journal}{Phys. Rev. X}} \textbf{\bibinfo{volume}{9}},
  \bibinfo{pages}{021060} (\bibinfo{year}{2019}).
\newblock \urlprefix\url{https://link.aps.org/doi/10.1103/PhysRevX.9.021060}.

\bibitem{Kumar2020}
\bibinfo{author}{Kumar, A.} \& \bibinfo{author}{Bechhoefer, J.}
\newblock \bibinfo{title}{Exponentially faster cooling in a colloidal system}.
\newblock \emph{\bibinfo{journal}{Nature}} \textbf{\bibinfo{volume}{584}},
  \bibinfo{pages}{64--68} (\bibinfo{year}{2020}).
\newblock \urlprefix\url{https://doi.org/10.1038/s41586-020-2560-x}.

\bibitem{Gal2020}
\bibinfo{author}{Gal, A.} \& \bibinfo{author}{Raz, O.}
\newblock \bibinfo{title}{Precooling strategy allows exponentially faster
  heating}.
\newblock \emph{\bibinfo{journal}{Phys. Rev. Lett.}}
  \textbf{\bibinfo{volume}{124}}, \bibinfo{pages}{060602}
  (\bibinfo{year}{2020}).
\newblock
  \urlprefix\url{https://link.aps.org/doi/10.1103/PhysRevLett.124.060602}.

\bibitem{Kumar2022}
\bibinfo{author}{Kumar, A.}, \bibinfo{author}{Ch{\'e}trite, R.} \&
  \bibinfo{author}{Bechhoefer, J.}
\newblock \bibinfo{title}{Anomalous heating in a colloidal system}.
\newblock \emph{\bibinfo{journal}{PNAS}} \textbf{\bibinfo{volume}{119}},
  \bibinfo{pages}{e2118484119} (\bibinfo{year}{2022}).
\newblock \urlprefix\url{https://doi.org/10.1073/pnas.2118484119}.

\bibitem{Teza2023}
\bibinfo{author}{Teza, G.}, \bibinfo{author}{Yaacoby, R.} \&
  \bibinfo{author}{Raz, O.}
\newblock \bibinfo{title}{Relaxation shortcuts through boundary coupling}.
\newblock \emph{\bibinfo{journal}{Phys. Rev. Lett.}}
  \textbf{\bibinfo{volume}{131}}, \bibinfo{pages}{017101}
  (\bibinfo{year}{2023}).
\newblock
  \urlprefix\url{https://link.aps.org/doi/10.1103/PhysRevLett.131.017101}.

\bibitem{Teza2025}
\bibinfo{author}{Teza, G.}, \bibinfo{author}{Bechhoefer, J.},
  \bibinfo{author}{Lasanta, A.}, \bibinfo{author}{Raz, O.} \&
  \bibinfo{author}{Vucelja, M.}
\newblock \bibinfo{title}{Speedups in nonequilibrium thermal relaxation: Mpemba
  and related effects}.
\newblock \emph{\bibinfo{journal}{Physics Reports}}
  \textbf{\bibinfo{volume}{1164}}, \bibinfo{pages}{1--97}
  (\bibinfo{year}{2026}).
\newblock
  \urlprefix\url{https://www.sciencedirect.com/science/article/pii/S0370157325002984}.

\bibitem{Nava2019}
\bibinfo{author}{Nava, A.} \& \bibinfo{author}{Fabrizio, M.}
\newblock \bibinfo{title}{Lindblad dissipative dynamics in the presence of
  phase coexistence}.
\newblock \emph{\bibinfo{journal}{Phys. Rev. B}}
  \textbf{\bibinfo{volume}{100}}, \bibinfo{pages}{125102}
  (\bibinfo{year}{2019}).
\newblock \urlprefix\url{https://link.aps.org/doi/10.1103/PhysRevB.100.125102}.

\bibitem{Carollo2021}
\bibinfo{author}{Carollo, F.}, \bibinfo{author}{Lasanta, A.} \&
  \bibinfo{author}{Lesanovsky, I.}
\newblock \bibinfo{title}{Exponentially accelerated approach to stationarity in
  markovian open quantum systems through the mpemba effect}.
\newblock \emph{\bibinfo{journal}{Phys. Rev. Lett.}}
  \textbf{\bibinfo{volume}{127}}, \bibinfo{pages}{060401}
  (\bibinfo{year}{2021}).
\newblock
  \urlprefix\url{https://link.aps.org/doi/10.1103/PhysRevLett.127.060401}.

\bibitem{Bao2022}
\bibinfo{author}{Bao, R.} \& \bibinfo{author}{Hou, Z.}
\newblock \bibinfo{title}{Accelerating quantum relaxation via temporary reset:
  A mpemba-inspired approach}.
\newblock \emph{\bibinfo{journal}{Phys. Rev. Lett.}}
  \textbf{\bibinfo{volume}{135}}, \bibinfo{pages}{150403}
  (\bibinfo{year}{2025}).
\newblock \urlprefix\url{https://link.aps.org/doi/10.1103/g94p-7421}.

\bibitem{Kochsiek2022}
\bibinfo{author}{Kochsiek, S.}, \bibinfo{author}{Carollo, F.} \&
  \bibinfo{author}{Lesanovsky, I.}
\newblock \bibinfo{title}{Accelerating the approach of dissipative quantum spin
  systems towards stationarity through global spin rotations}.
\newblock \emph{\bibinfo{journal}{Phys. Rev. A}}
  \textbf{\bibinfo{volume}{106}}, \bibinfo{pages}{012207}
  (\bibinfo{year}{2022}).
\newblock \urlprefix\url{https://link.aps.org/doi/10.1103/PhysRevA.106.012207}.

\bibitem{Ivander2023}
\bibinfo{author}{Ivander, F.}, \bibinfo{author}{Anto-Sztrikacs, N.} \&
  \bibinfo{author}{Segal, D.}
\newblock \bibinfo{title}{Hyperacceleration of quantum thermalization dynamics
  by bypassing long-lived coherences: An analytical treatment}.
\newblock \emph{\bibinfo{journal}{Phys. Rev. E}}
  \textbf{\bibinfo{volume}{108}}, \bibinfo{pages}{014130}
  (\bibinfo{year}{2023}).
\newblock \urlprefix\url{https://link.aps.org/doi/10.1103/PhysRevE.108.014130}.

\bibitem{Wang2024}
\bibinfo{author}{Wang, X.} \& \bibinfo{author}{Wang, J.}
\newblock \bibinfo{title}{Mpemba effects in nonequilibrium open quantum
  systems}.
\newblock \emph{\bibinfo{journal}{Phys. Rev. Res.}}
  \textbf{\bibinfo{volume}{6}}, \bibinfo{pages}{033330} (\bibinfo{year}{2024}).
\newblock
  \urlprefix\url{https://link.aps.org/doi/10.1103/PhysRevResearch.6.033330}.

\bibitem{Moroder2024}
\bibinfo{author}{Moroder, M.}, \bibinfo{author}{Culhane, O.},
  \bibinfo{author}{Zawadzki, K.} \& \bibinfo{author}{Goold, J.}
\newblock \bibinfo{title}{Thermodynamics of the quantum mpemba effect}.
\newblock \emph{\bibinfo{journal}{Phys. Rev. Lett.}}
  \textbf{\bibinfo{volume}{133}}, \bibinfo{pages}{140404}
  (\bibinfo{year}{2024}).
\newblock
  \urlprefix\url{https://link.aps.org/doi/10.1103/PhysRevLett.133.140404}.

\bibitem{Strachan2024}
\bibinfo{author}{Strachan, D.~J.}, \bibinfo{author}{Purkayastha, A.} \&
  \bibinfo{author}{Clark, S.~R.}
\newblock \bibinfo{title}{Non-markovian quantum mpemba effect}.
\newblock \emph{\bibinfo{journal}{Phys. Rev. Lett.}}
  \textbf{\bibinfo{volume}{134}}, \bibinfo{pages}{220403}
  (\bibinfo{year}{2025}).
\newblock
  \urlprefix\url{https://link.aps.org/doi/10.1103/PhysRevLett.134.220403}.

\bibitem{Xu2025}
\bibinfo{author}{Xu, M.}, \bibinfo{author}{Wei, Z.}, \bibinfo{author}{Jiang,
  X.-P.} \& \bibinfo{author}{Pan, L.}
\newblock \bibinfo{title}{Expedited thermalization dynamics in incommensurate
  systems} (\bibinfo{year}{2025}).
\newblock \urlprefix\url{https://arxiv.org/abs/2505.03645}.
\newblock
  \bibinfo{eprint}{{\href{https://arxiv.org/abs/2505.03645}{{arXiv:2505.03645}}}}.

\bibitem{Longhi2025}
\bibinfo{author}{Longhi, S.}
\newblock \bibinfo{title}{Mpemba effect and super-accelerated thermalization in
  the damped quantum harmonic oscillator}.
\newblock \emph{\bibinfo{journal}{{Quantum}}} \textbf{\bibinfo{volume}{9}},
  \bibinfo{pages}{1677} (\bibinfo{year}{2025}).
\newblock \urlprefix\url{https://doi.org/10.22331/q-2025-03-26-1677}.

\bibitem{Ares2023Nat}
\bibinfo{author}{Ares, F.}, \bibinfo{author}{Murciano, S.} \&
  \bibinfo{author}{Calabrese, P.}
\newblock \bibinfo{title}{Entanglement asymmetry as a probe of symmetry
  breaking}.
\newblock \emph{\bibinfo{journal}{Nat. Commun.}} \textbf{\bibinfo{volume}{14}},
  \bibinfo{pages}{2036} (\bibinfo{year}{2023}).
\newblock \urlprefix\url{https://doi.org/10.1038/s41467-023-37747-8}.

\bibitem{Ares2025}
\bibinfo{author}{Ares, F.}, \bibinfo{author}{Calabrese, P.} \&
  \bibinfo{author}{Murciano, S.}
\newblock \bibinfo{title}{The quantum mpemba effects}.
\newblock \emph{\bibinfo{journal}{Nat. Rev. Phys.}}  (\bibinfo{year}{2025}).
\newblock \urlprefix\url{http://dx.doi.org/10.1038/s42254-025-00838-0}.

\bibitem{Rylands2024}
\bibinfo{author}{Rylands, C.} \emph{et~al.}
\newblock \bibinfo{title}{Microscopic origin of the quantum mpemba effect in
  integrable systems}.
\newblock \emph{\bibinfo{journal}{Phys. Rev. Lett.}}
  \textbf{\bibinfo{volume}{133}}, \bibinfo{pages}{010401}
  (\bibinfo{year}{2024}).
\newblock
  \urlprefix\url{https://link.aps.org/doi/10.1103/PhysRevLett.133.010401}.

\bibitem{Turkeshi2024}
\bibinfo{author}{Turkeshi, X.}, \bibinfo{author}{Calabrese, P.} \&
  \bibinfo{author}{{De Luca}, A.}
\newblock \bibinfo{title}{Quantum mpemba effect in random circuits}.
\newblock \emph{\bibinfo{journal}{Phys. Rev. Lett.}}  (\bibinfo{year}{2025}).
\newblock \urlprefix\url{http://dx.doi.org/10.1103/5d6p-8d1b}.

\bibitem{Liu2024}
\bibinfo{author}{Liu, S.}, \bibinfo{author}{Zhang, H.-K.},
  \bibinfo{author}{Yin, S.} \& \bibinfo{author}{Zhang, S.-X.}
\newblock \bibinfo{title}{Symmetry restoration and quantum mpemba effect in
  symmetric random circuits}.
\newblock \emph{\bibinfo{journal}{Phys. Rev. Lett.}}
  \textbf{\bibinfo{volume}{133}}, \bibinfo{pages}{140405}
  (\bibinfo{year}{2024}).
\newblock
  \urlprefix\url{https://link.aps.org/doi/10.1103/PhysRevLett.133.140405}.

\bibitem{Li2025}
\bibinfo{author}{Li, Y.}, \bibinfo{author}{Li, W.} \& \bibinfo{author}{Li, X.}
\newblock \bibinfo{title}{Ergotropic mpemba effect in non-markovian quantum
  systems}.
\newblock \emph{\bibinfo{journal}{Phys. Rev. A}}
  \textbf{\bibinfo{volume}{112}}, \bibinfo{pages}{032209}
  (\bibinfo{year}{2025}).
\newblock \urlprefix\url{https://link.aps.org/doi/10.1103/5xrr-x2rm}.

\bibitem{Summer2026}
\bibinfo{author}{Summer, A.} \emph{et~al.}
\newblock \bibinfo{title}{Resource-theoretical unification of mpemba effects:
  Classical and quantum}.
\newblock \emph{\bibinfo{journal}{Phys. Rev. X}} \textbf{\bibinfo{volume}{16}},
  \bibinfo{pages}{011065} (\bibinfo{year}{2026}).
\newblock \urlprefix\url{https://link.aps.org/doi/10.1103/rbt4-psfd}.

\bibitem{Beato2026}
\bibinfo{author}{Beato, N.} \& \bibinfo{author}{Teza, G.}
\newblock \bibinfo{title}{Relaxation control of open quantum systems}.
\newblock \emph{\bibinfo{journal}{Phys. Rev. Lett.}}
  \textbf{\bibinfo{volume}{136}}, \bibinfo{pages}{070401}
  (\bibinfo{year}{2026}).
\newblock \urlprefix\url{https://link.aps.org/doi/10.1103/4frd-ck2z}.

\bibitem{Aristotle}
\bibinfo{author}{Aristotle} \& \bibinfo{author}{Lee, P. H.~D.}
\newblock \emph{\bibinfo{title}{Book I Chapter XII}}, \bibinfo{pages}{79–86}
  (\bibinfo{publisher}{Harvard University Press}, \bibinfo{year}{1952}).
\newblock \urlprefix\url{https://www.hup.harvard.edu/books/9780674994362}.

\bibitem{Mpemba1969}
\bibinfo{author}{Mpemba, E.~B.} \& \bibinfo{author}{Osborne, D.~G.}
\newblock \bibinfo{title}{Cool?}
\newblock \emph{\bibinfo{journal}{Phys. Educ.}} \textbf{\bibinfo{volume}{4}},
  \bibinfo{pages}{172} (\bibinfo{year}{1969}).
\newblock \urlprefix\url{https://dx.doi.org/10.1088/0031-9120/4/3/312}.

\bibitem{Ze-Zhou2025}
\bibinfo{author}{Zhang, Z.-Z.}, \bibinfo{author}{Luo, H.-G.} \&
  \bibinfo{author}{Wu, W.}
\newblock \bibinfo{title}{Quantum mpemba effect induced by non-markovian
  exceptional points} (\bibinfo{year}{2025}).
\newblock \urlprefix\url{https://arxiv.org/abs/2511.13173}.
\newblock
  \bibinfo{eprint}{{\href{https://arxiv.org/abs/2511.13173}{{arXiv:2511.13173}}}}.

\bibitem{Aharony2024}
\bibinfo{author}{Aharony~Shapira, S.} \emph{et~al.}
\newblock \bibinfo{title}{Inverse mpemba effect demonstrated on a single
  trapped ion qubit}.
\newblock \emph{\bibinfo{journal}{Phys. Rev. Lett.}}
  \textbf{\bibinfo{volume}{133}}, \bibinfo{pages}{010403}
  (\bibinfo{year}{2024}).
\newblock
  \urlprefix\url{https://link.aps.org/doi/10.1103/PhysRevLett.133.010403}.

\bibitem{Joshi2024}
\bibinfo{author}{Joshi, L.~K.} \emph{et~al.}
\newblock \bibinfo{title}{Observing the quantum mpemba effect in quantum
  simulations}.
\newblock \emph{\bibinfo{journal}{Phys. Rev. Lett.}}
  \textbf{\bibinfo{volume}{133}}, \bibinfo{pages}{010402}
  (\bibinfo{year}{2024}).
\newblock
  \urlprefix\url{https://link.aps.org/doi/10.1103/PhysRevLett.133.010402}.

\bibitem{Zhang2025}
\bibinfo{author}{Zhang, J.} \emph{et~al.}
\newblock \bibinfo{title}{Observation of quantum strong mpemba effect}.
\newblock \emph{\bibinfo{journal}{Nat. Comm.}} \textbf{\bibinfo{volume}{16}},
  \bibinfo{pages}{301} (\bibinfo{year}{2025}).
\newblock \urlprefix\url{https://www.nature.com/articles/s41467-024-54303-0}.

\bibitem{Nava2024}
\bibinfo{author}{Nava, A.} \& \bibinfo{author}{Egger, R.}
\newblock \bibinfo{title}{Mpemba effects in open nonequilibrium quantum
  systems}.
\newblock \emph{\bibinfo{journal}{Phys. Rev. Lett.}}
  \textbf{\bibinfo{volume}{133}}, \bibinfo{pages}{136302}
  (\bibinfo{year}{2024}).
\newblock
  \urlprefix\url{https://link.aps.org/doi/10.1103/PhysRevLett.133.136302}.

\bibitem{WangWang2024}
\bibinfo{author}{Wang, X.} \& \bibinfo{author}{Wang, J.}
\newblock \bibinfo{title}{Mpemba effects in nonequilibrium open quantum
  systems}.
\newblock \emph{\bibinfo{journal}{Phys. Rev. Res.}}
  \textbf{\bibinfo{volume}{6}}, \bibinfo{pages}{033330} (\bibinfo{year}{2024}).
\newblock
  \urlprefix\url{https://link.aps.org/doi/10.1103/PhysRevResearch.6.033330}.

\bibitem{Westhoff2025}
\bibinfo{author}{Westhoff, P.}, \bibinfo{author}{Paeckel, S.} \&
  \bibinfo{author}{Moroder, M.}
\newblock \bibinfo{title}{Fast and direct preparation of a genuine lattice
  bose-einstein condensate via the quantum mpemba effect}.
\newblock \emph{\bibinfo{journal}{Phys. Rev. A}}
  \textbf{\bibinfo{volume}{112}}, \bibinfo{pages}{L061304}
  (\bibinfo{year}{2025}).
\newblock \urlprefix\url{https://link.aps.org/doi/10.1103/nm4b-9w5c}.

\bibitem{Mondal2026}
\bibinfo{author}{Mondal, S.} \& \bibinfo{author}{Sen, U.}
\newblock \bibinfo{title}{Mpemba effect in self-contained quantum
  refrigerators: Accelerated cooling}.
\newblock \emph{\bibinfo{journal}{Physical Review A}}
  \textbf{\bibinfo{volume}{113}} (\bibinfo{year}{2026}).
\newblock \urlprefix\url{http://dx.doi.org/10.1103/8fp4-15mq}.

\bibitem{Medina2024}
\bibinfo{author}{Medina, I.}, \bibinfo{author}{Culhane, O.},
  \bibinfo{author}{Binder, F.~C.}, \bibinfo{author}{Landi, G.~T.} \&
  \bibinfo{author}{Goold, J.}
\newblock \bibinfo{title}{Anomalous discharging of quantum batteries: The
  ergotropic mpemba effect}.
\newblock \emph{\bibinfo{journal}{Phys. Rev. Lett.}}
  \textbf{\bibinfo{volume}{134}}, \bibinfo{pages}{220402}
  (\bibinfo{year}{2025}).
\newblock
  \urlprefix\url{https://link.aps.org/doi/10.1103/PhysRevLett.134.220402}.

\bibitem{Lejeune2026}
\bibinfo{author}{Lejeune, T.} \emph{et~al.}
\newblock \bibinfo{title}{Accelerating qubit reset through the mpemba effect}
  (\bibinfo{year}{2026}).
\newblock \urlprefix\url{https://arxiv.org/abs/2602.03765}.
\newblock
  \bibinfo{eprint}{{\href{https://arxiv.org/abs/2602.03765}{{arXiv:2602.03765}}}}.

\bibitem{Whitelam2025_clock}
\bibinfo{author}{Whitelam, S.}
\newblock \bibinfo{title}{Increasing the clock speed of a thermodynamic
  computer by adding noise}.
\newblock \emph{\bibinfo{journal}{npj Unconventional Computing}}
  \textbf{\bibinfo{volume}{2}}, \bibinfo{pages}{24} (\bibinfo{year}{2025}).
\newblock \urlprefix\url{https://www.nature.com/articles/s44335-025-00038-0}.

\bibitem{Gardiner2004}
\bibinfo{author}{Gardiner, C.}
\newblock \emph{\bibinfo{title}{Handbook of Stochastic Methods for Physics,
  Chemistry, and the Natural Sciences}} Springer complexity
  (\bibinfo{publisher}{Springer}, \bibinfo{year}{2004}).
\newblock \urlprefix\url{https://books.google.it/books?id=wLm7QgAACAAJ}.

\bibitem{LiberzonBrockett2000}
\bibinfo{author}{Liberzon, D.} \& \bibinfo{author}{Brockett, R.~W.}
\newblock \bibinfo{title}{Spectral analysis of {Fokker--Planck} and related
  operators arising from linear stochastic differential equations}.
\newblock \emph{\bibinfo{journal}{SIAM Journal on Control and Optimization}}
  \textbf{\bibinfo{volume}{38}}, \bibinfo{pages}{1453--1467}
  (\bibinfo{year}{2000}).
\newblock \urlprefix\url{https://doi.org/10.1137/S0363012998338193}.

\bibitem{Lanczos1950}
\bibinfo{author}{Lanczos, C.}
\newblock \bibinfo{title}{{An iteration method for the solution of the
  eigenvalue problem of linear differential and integral operators}}.
\newblock \emph{\bibinfo{journal}{J. Res. Natl. Bur. Stand. B}}
  \textbf{\bibinfo{volume}{45}}, \bibinfo{pages}{255--282}
  (\bibinfo{year}{1950}).

\bibitem{Saad2011}
\bibinfo{author}{Saad, Y.}
\newblock \emph{\bibinfo{title}{Numerical Methods for Large Eigenvalue
  Problems}} \bibinfo{edition}{2} edn (\bibinfo{publisher}{Society for
  Industrial and Applied Mathematics}, \bibinfo{address}{Philadelphia, PA},
  \bibinfo{year}{2011}).
\newblock
  \urlprefix\url{https://epubs.siam.org/doi/book/10.1137/1.9781611970739}.

\bibitem{Marcenko_1967}
\bibinfo{author}{Marčenko, V.~A.} \& \bibinfo{author}{Pastur, L.~A.}
\newblock \bibinfo{title}{Distribution of eigenvalues for some sets of random
  matrices}.
\newblock \emph{\bibinfo{journal}{Mathematics of the USSR-Sbornik}}
  \textbf{\bibinfo{volume}{1}}, \bibinfo{pages}{457} (\bibinfo{year}{1967}).
\newblock \urlprefix\url{https://doi.org/10.1070/SM1967v001n04ABEH001994}.

\bibitem{BaiSilverstein2010}
\bibinfo{author}{Bai, Z.~D.} \& \bibinfo{author}{Silverstein, J.~W.}
\newblock \emph{\bibinfo{title}{Spectral Analysis of Large Dimensional Random
  Matrices}}  (\bibinfo{publisher}{Springer}, \bibinfo{year}{2010}).
\newblock
  \urlprefix\url{https://link.springer.com/book/10.1007/978-1-4419-0661-8}.

\bibitem{RasmussenWilliams2006}
\bibinfo{author}{Rasmussen, C.~E.} \& \bibinfo{author}{Williams, C. K.~I.}
\newblock \emph{\bibinfo{title}{Gaussian Processes for Machine Learning}}
  (\bibinfo{publisher}{The MIT Press}, \bibinfo{year}{2005}).
\newblock \urlprefix\url{https://doi.org/10.7551/mitpress/3206.001.0001}.

\bibitem{Hoerl01021970}
\bibinfo{author}{Hoerl, A.~E.} \& \bibinfo{author}{Kennard, R.~W.}
\newblock \bibinfo{title}{Ridge regression: Biased estimation for nonorthogonal
  problems}.
\newblock \emph{\bibinfo{journal}{Technometrics}}
  \textbf{\bibinfo{volume}{12}}, \bibinfo{pages}{55--67}
  (\bibinfo{year}{1970}).
\newblock \urlprefix\url{https://doi.org/10.1080/00401706.1970.10488634}.

\bibitem{Crooks1999}
\bibinfo{author}{Crooks, G.~E.}
\newblock \bibinfo{title}{Entropy production fluctuation theorem and the
  nonequilibrium work relation for free energy differences}.
\newblock \emph{\bibinfo{journal}{Phys. Rev. E}} \textbf{\bibinfo{volume}{60}},
  \bibinfo{pages}{2721--2726} (\bibinfo{year}{1999}).
\newblock \urlprefix\url{https://link.aps.org/doi/10.1103/PhysRevE.60.2721}.

\bibitem{Jarzynski1997}
\bibinfo{author}{Jarzynski, C.}
\newblock \bibinfo{title}{Nonequilibrium equality for free energy differences}.
\newblock \emph{\bibinfo{journal}{Phys. Rev. Lett.}}
  \textbf{\bibinfo{volume}{78}}, \bibinfo{pages}{2690--2693}
  (\bibinfo{year}{1997}).
\newblock \urlprefix\url{https://link.aps.org/doi/10.1103/PhysRevLett.78.2690}.

\bibitem{Bennett1976}
\bibinfo{author}{Bennett, C.~H.}
\newblock \bibinfo{title}{Efficient estimation of free energy differences from
  monte carlo data}.
\newblock \emph{\bibinfo{journal}{Journal of Computational Physics}}
  \textbf{\bibinfo{volume}{22}}, \bibinfo{pages}{245--268}
  (\bibinfo{year}{1976}).
\newblock
  \urlprefix\url{https://www.sciencedirect.com/science/article/pii/0021999176900784}.

\bibitem{Basak2026}
\bibinfo{author}{Basak, P.}, \bibinfo{author}{Whitelam, S.} \&
  \bibinfo{author}{Bechhoefer, J.}
\newblock \bibinfo{title}{Adding noise and scaling forces to speed up the
  langevin clock} (\bibinfo{year}{2026}).
\newblock \urlprefix\url{https://arxiv.org/abs/2603.03620}.
\newblock
  \bibinfo{eprint}{{\href{https://arxiv.org/abs/2603.03620}{{arXiv:2603.03620}}}}.

\bibitem{Freitas2021}
\bibinfo{author}{Freitas, N.}, \bibinfo{author}{Delvenne, J.-C.} \&
  \bibinfo{author}{Esposito, M.}
\newblock \bibinfo{title}{Stochastic thermodynamics of nonlinear electronic
  circuits: A realistic framework for computing around $kt$}.
\newblock \emph{\bibinfo{journal}{Phys. Rev. X}} \textbf{\bibinfo{volume}{11}},
  \bibinfo{pages}{031064} (\bibinfo{year}{2021}).
\newblock \urlprefix\url{https://link.aps.org/doi/10.1103/PhysRevX.11.031064}.

\bibitem{Lipka-Bartosik2024}
\bibinfo{author}{Lipka-Bartosik, P.}, \bibinfo{author}{Perarnau-Llobet, M.} \&
  \bibinfo{author}{Brunner, N.}
\newblock \bibinfo{title}{Thermodynamic computing via autonomous quantum
  thermal machines}.
\newblock \emph{\bibinfo{journal}{Science Advances}}
  \textbf{\bibinfo{volume}{10}}, \bibinfo{pages}{eadm8792}
  (\bibinfo{year}{2024}).
\newblock
  \urlprefix\url{https://www.science.org/doi/abs/10.1126/sciadv.adm8792}.

\end{thebibliography}


\begin{thebibliography}{1}
\expandafter\ifx\csname url\endcsname\relax
  \def\url#1{\burl{#1}}\fi
\expandafter\ifx\csname urlprefix\endcsname\relax\def\urlprefix{URL }\fi
\providecommand{\bibinfo}[2]{#2}
\providecommand{\eprint}[2][]{\url{#2}}
\providecommand{\doi}[1]{\url{https://doi.org/#1}}
\bibcommenthead

\bibitem{Gardiner2004}
\bibinfo{author}{Gardiner, C.}
\newblock \emph{\bibinfo{title}{Handbook of Stochastic Methods for Physics,
  Chemistry, and the Natural Sciences}} Springer complexity
  (\bibinfo{publisher}{Springer}, \bibinfo{year}{2004}).
\newblock \urlprefix\url{https://books.google.it/books?id=wLm7QgAACAAJ}.

\bibitem{Melanson2023}
\bibinfo{author}{Melanson, D.} \emph{et~al.}
\newblock \bibinfo{title}{Thermodynamic computing system for {AI}
  applications}.
\newblock \emph{\bibinfo{journal}{Nature Communications}}
  \textbf{\bibinfo{volume}{16}}, \bibinfo{pages}{3757} (\bibinfo{year}{2025}).
\newblock
  \urlprefix\url{https://www.nature.com/articles/s41467-025-59011-x#citeas}.

\bibitem{Saad2011}
\bibinfo{author}{Saad, Y.}
\newblock \emph{\bibinfo{title}{Numerical Methods for Large Eigenvalue
  Problems}} \bibinfo{edition}{2} edn (\bibinfo{publisher}{Society for
  Industrial and Applied Mathematics}, \bibinfo{address}{Philadelphia, PA},
  \bibinfo{year}{2011}).
\newblock
  \urlprefix\url{https://epubs.siam.org/doi/book/10.1137/1.9781611970739}.

\bibitem{RasmussenWilliams2006}
\bibinfo{author}{Rasmussen, C.~E.} \& \bibinfo{author}{Williams, C. K.~I.}
\newblock \emph{\bibinfo{title}{Gaussian Processes for Machine Learning}}
  (\bibinfo{publisher}{The MIT Press}, \bibinfo{year}{2005}).
\newblock \urlprefix\url{https://doi.org/10.7551/mitpress/3206.001.0001}.

\bibitem{Zhu1998}
\bibinfo{author}{Zhu, H.}, \bibinfo{author}{Williams, C. K.~I.},
  \bibinfo{author}{Rohwer, R.}, \bibinfo{author}{Morciniec, M.} \&
  \bibinfo{author}{Hammel, M.}
\newblock \bibinfo{title}{Gaussian regression and optimal finite dimensional
  linear models}.
\newblock \bibinfo{type}{Tech. Rep.} \bibinfo{number}{NCRG/97/011},
  \bibinfo{institution}{Aston University} (\bibinfo{year}{1997}).
\newblock
  \urlprefix\url{https://www.research.ed.ac.uk/en/publications/gaussian-regression-and-optimal-finite-dimensional-linear-models}.

\bibitem{Belkin2018}
\bibinfo{author}{Belkin, M.}
\newblock \bibinfo{title}{Approximation beats concentration? an approximation
  view on inference with smooth radial kernels} (\bibinfo{year}{2018}).
\newblock \urlprefix\url{https://arxiv.org/abs/1801.03437}.
\newblock
  \bibinfo{eprint}{{\href{https://arxiv.org/abs/1801.03437}{{arXiv:1801.03437}}}}.

\end{thebibliography}

\end{document}


\newcommand{\papertitle}{%
Digitally Optimized Initializations for Fast Thermodynamic Computing%
}

\title{%
\centering Supplementary Information for\\[0.4em]
\textit{\papertitle}%
}

\author*[1]{\fnm{Mattia} \sur{Moroder}}\email{moroderm@tcd.ie}
\author[1]{\fnm{Felix C.} \sur{Binder}}
\author[1]{\fnm{John} \sur{Goold}}

\affil[1]{\orgdiv{School of Physics}, \orgname{Trinity College Dublin},
  \orgaddress{\street{College Green}, \city{Dublin 2},
  \postcode{D02K8N4}, \country{Ireland}}}

\maketitle
\setcounter{figure}{0}
\renewcommand{\thefigure}{S\arabic{figure}}

\appendix
\suppnote{Derivation of the Lyapunov equation for the covariance}
\label{app:lyapunov}

We start from the overdamped Langevin dynamics
\begin{equation}
\dot x_i(t)
=
-\mu\,\frac{\partial V(\textbf{x})}{\partial x_i}
+
\sqrt{2\mu k_B T}\,\eta_i(t),
\end{equation}
where $\textbf{x}(t)\in\mathbb{R}^d$, $\mu$ is the mobility, and $\eta_i(t)$ are independent
Gaussian white noises with
$\langle \eta_i(t)\eta_j(t')\rangle=\delta_{ij}\delta(t-t')$.
For a quadratic potential $V(\textbf{x})=\tfrac12 \textbf{x}^{T}\mathbf{A}\textbf{x}$, with $\mathbf{A}$
symmetric and positive definite, the Langevin equation can be written in It\^o
form as~\cite{Gardiner2004}
\begin{equation}
d\textbf{x}(t) = -\mu \mathbf{A}\textbf{x}(t)\,dt + \sqrt{2\mu k_B T}\,d\textbf{W}(t),
\end{equation}
where $\textbf{W}(t)$ is a $d$-dimensional Wiener process satisfying
$\langle dW\,dW^{T}\rangle=\mathbb{1}\,dt$.

Defining the covariance matrix as
$\mathbf{\Sigma}(t)\equiv\langle \mathbf{x}(t)\mathbf{x}(t)^{T}\rangle$, and applying It\^o’s product rule to $\mathbf{x} \mathbf{x}^{T}$, one finds
$d(\mathbf{x}\mathbf{x}^{T})
=
(d\mathbf{x})\,\mathbf{x}^{T}
+
\mathbf{x}\,(d\mathbf{x})^{T}
+
(d\mathbf{x})(d\mathbf{x})^{T}
$.
Using the symmetry of $\mathbf{A}$, for the Langevin equation this yields
$(d\mathbf{x})\,\mathbf{x}^{T}
+
\mathbf{x}\,(d\mathbf{x})^{T}
=
-\mu\bigl(\mathbf{A}\mathbf{x}\mathbf{x}^{T}
+
\mathbf{x}\mathbf{x}^{T}\mathbf{A}\bigr)\,dt$ and $(d\mathbf{x})(d\mathbf{x})^{T}
=
2\mu k_B T\,\mathbb{1}\,dt
$.
Finally, taking the average and using $\mathbf{\Sigma}(t)=\langle \mathbf{x}\mathbf{x}^{T}\rangle$ then gives the
closed evolution equation
\begin{equation}
\dot{\mathbf{\Sigma}}(t)
=
-\mu\bigl(\mathbf{A}\mathbf{\Sigma}(t)+\mathbf{\Sigma}(t)\mathbf{A}\bigr)
+2\mu k_B T\,\mathbb{1}.
\end{equation}
This is the continuous-time Lyapunov equation governing the covariance dynamics
of the overdamped Langevin process.

\suppnote{\updated{Extension to underdamped dynamics}}
\label{app:underdamped}

Some experimental implementations of thermodynamic computing, in
particular RLC-based circuits, can operate outside the overdamped regime.
We therefore outline how the optimized-initialization protocol extends to
underdamped Langevin dynamics. We first consider the direct underdamped
analogue of the overdamped model used in the main text, in which the
programmable matrix enters the potential, or stiffness, sector. Thus
\(x\in\mathbb{R}^d\) evolves in the quadratic potential
\(U(x)=\frac12 x^T \mathbf A x\), with \(\mathbf A\succ0\), friction \(\gamma\),
inverse temperature \(\beta\), and scalar mass \(m\). The corresponding
phase-space variable is \(\mathbf{z}=(\mathbf{x},\mathbf{p})\). In this stiffness-sector encoding, the
stationary position covariance is \(\beta^{-1}\mathbf{A}^{-1}\). The complementary
capacitance, or mass-sector, encoding used in the RLC implementation of Ref.~\cite{Melanson2023} is discussed at the end.

The pair $\mathbf{z}=(\bm{x},\bm{p})$ is an
Ornstein-Uhlenbeck process $\ud \mathbf{z}=\mathbf{F}\mathbf{z}\,\ud t+\sqrt{2\mathbf{D}}\,\ud \mathbf{W}$ with
\begin{equation}
\mathbf{F}=\begin{pmatrix}0 & m^{-1}\mathbb{1}\\[1pt] -A & -\tfrac{\gamma}{m}\mathbb{1}\end{pmatrix},
\qquad
\mathbf{D}=\begin{pmatrix}0&0\\[1pt] 0&\gamma\beta^{-1}\mathbb{1}\end{pmatrix}.
\end{equation}
The phase-space covariance obeys $\dot{\mathbf{\Sigma}}=\mathbf{F}\mathbf{\Sigma}+\mathbf{\Sigma} \mathbf{F}^\top+2\mathbf{D}$, with
equipartition fixed point $\Sss=\beta^{-1}\mathrm{diag}(\mathbf{A}^{-1},\,m\,\mathbb{1})$,
so the readout of $\mathbf{A}^{-1}$ from the position block is unchanged. The
deviation $\dS=\mathbf{\Sigma}-\Sss$ obeys the homogeneous equation, solved exactly by
\begin{equation}
\dS(t)=e^{\mathbf{F}t}\,\dS(0)\,e^{\mathbf{F}^T t},
\label{eq:D-homo}
\end{equation}
so the relaxation, and hence the Mpemba construction, is governed by
the spectrum of $\mathbf{F}$, as it was by $-\mathbf{A}/\gamma$ in the overdamped case.
 
Writing $\mathbf{A}=\sum_k\lambda_k\mathbf{u}_k\mathbf{u}_k^T$,
$\lambda_1\le\dots\le\lambda_d$, and noting that $\mathbf{F}$ leaves each
two-dimensional subspace $\mathrm{span}\{(\mathbf{u}_k,0),(0,\mathbf{u}_k)\}$
invariant, the drift matrix decomposes into $d$ independent $2\times2$ blocks
$\mathbf{F}_k=\bigl(\begin{smallmatrix}0&1/m\\-\lambda_k&-\gamma/m\end{smallmatrix}\bigr)$.
The eigenvalues $\mu_\pm(\lambda_k)$ of $\mathbf{F}_k$ are the roots of its
characteristic polynomial $\det(\mathbf{F}_k-\mu\mathbb{1})=m^{-1}(m\mu^2+\gamma\mu+\lambda_k)=0$,
\begin{equation}
\mu_\pm(\lambda_k)=-\tilde\gamma\pm\sqrt{\tilde\gamma^{\,2}-\lambda_k/m},
\qquad \tilde\gamma\equiv\frac{\gamma}{2m}.
\end{equation}
This introduces the critical eigenvalue
\begin{equation}
\lambda^\star\equiv\frac{\gamma^2}{4m},
\label{eq:D-astar}
\end{equation}
separating per-mode overdamped ($\lambda_k<\lambda^\star$) from underdamped
behavior. The per-mode envelope rate $r(\lambda)\equiv-\mathrm{Re}\,\mu_+(\lambda)$ is
\begin{equation}
r(\lambda)=
\begin{cases}
\gt\bigl(1-\sqrt{1-\lambda/\lambda^\star}\bigr), & \lambda<\lambda^\star,\\[3pt]
\gt=\gamma/2m, & \lambda\ge\lambda^\star,
\end{cases}
\label{eq:D-rate}
\end{equation}
underdamped modes oscillating in addition at
$\Omega_k=\sqrt{\lambda_k/m-\gt^{\,2}}$. With $\mathbf{v}_j$, $\mathbf{w}_j$ the right and left
eigenvectors of $\mathbf{F}$ ($\mathbf{w}_j^\top \mathbf{v}_k=\delta_{jk}$), Eq.~\eqref{eq:D-homo}
expands as
\begin{equation}
\dS(t)=\textstyle\sum_{jk}c_{jk}\,\mathbf{v}_j\mathbf{v}_k^\top e^{(\mu_j+\mu_k)t},
\quad c_{jk}=\mathbf{w}_j^\top\dS(0)\,\mathbf{w}_k,
\label{eq:D-expansion}
\end{equation}
so the covariance relaxes at rates $r(\lambda_j)+r(\lambda_k)$, the slowest
term decaying as $e^{-2r(\lambda_1)t}$. Note that $r(\lambda)$ increases
monotonically for $\lambda<\lambda^\star$ and is then \emph{flat}:
anomalously slow modes exist only below $\lambda^\star$. In the overdamped
limit $m\to0$ (equivalently $\lambda_d\ll\lambda^\star$, per-mode quality
factor $Q_k=\sqrt{m\lambda_k}/\gamma\ll1/2$) one has
$r(\lambda)\to\lambda/\gamma$, recovering the rate spectrum of the main text,
while the fast branch $\mu_-\to-\gamma/m$ is the adiabatically eliminated
momentum relaxation.
 
Up to normalization, the left eigenvectors of $\mathbf{F}_k$ are
$\mathbf{w}_k^{(\pm)}=(\lambda_k \mathbf{u}_k,\,-\mu_\pm(\lambda_k)\mathbf{u}_k)$, so the exact
generalization of the initialization condition (remove the overlap of
$\dS(0)$ with the $K$ slowest modes) reads $\dS(0)\,\mathbf{w}_k^{(+)}=0$, i.e.
\begin{equation}
\begin{aligned}
\lambda_k\,\dS_{xx}\mathbf{u}_k&=\mu_+(\lambda_k)\,\dS_{xp}\mathbf{u}_k,\\
\lambda_k\,\dS_{px}\mathbf{u}_k&=\mu_+(\lambda_k)\,\dS_{pp}\mathbf{u}_k,
\end{aligned}
\qquad k\le K;
\label{eq:D-mpemba}
\end{equation}
by symmetry of $\dS(0)$ this removes every term in
Eq.~\eqref{eq:D-expansion} carrying a slow index. A sufficient realization is
to match all three covariance blocks to their stationary values on the
Lanczos subspace $\mathrm{span}\{\mathbf{u}_1,\dots,\mathbf{u}_K\}$:
$\mathbf{\Sigma}_{xx}(0)\mathbf{u}_k=\beta^{-1}\lambda_k^{-1}\mathbf{u}_k$,
$\mathbf{\Sigma}_{pp}(0)\mathbf{u}_k=\beta^{-1}m\,\mathbf{u}_k$, $\mathbf{\Sigma}_{xp}(0)\mathbf{u}_k=\mathbf{\Sigma}_{px}(0)\mathbf{u}_k=0$, using exactly
the same Lanczos data $(\lambda_k,\mathbf{u}_k)$ as the overdamped protocol. For the
stiffness-sector encoding the stationary momentum covariance
$\beta^{-1}m\,\mathbb{1}$ is isotropic and independent of $\mathbf{A}$; if the hardware
initializes the momentum sector in its thermal marginal, uncorrelated with
positions, then $\dS_{pp}(0)=\dS_{xp}(0)=0$ identically and
Eq.~\eqref{eq:D-mpemba} collapses to
$\dS_{xx}(0)\,\mathbf{u}_k=0$, $k\le K$, i.e. exactly the overdamped condition.
Under this implementation-dependent assumption the digital preprocessing
stage is therefore unchanged.
 
After projecting out the $K$ slowest modes, the residual slowest covariance
rate is $2r(\lambda_{K+1})$, giving the speedup
$S_{\rm ud}(K)=r(\lambda_{K+1})/r(\lambda_1)$ with $r$ from
Eq.~\eqref{eq:D-rate}; when both $\lambda_1,\lambda_{K+1}<\lambda^\star$,
\begin{equation}
S_{\rm ud}(K)
=\frac{1-\sqrt{1-\lambda_{K+1}/\lambda^\star}}{1-\sqrt{1-\lambda_1/\lambda^\star}}
\;\xrightarrow[\lambda_1,\lambda_{K+1}\ll\lambda^\star]{}\;
\frac{\lambda_{K+1}}{\lambda_1},
\end{equation}
recovering the overdamped result as a limit. Because $r$ saturates at $\gt$,
the projection stops improving the asymptotic decay exponent once
$\lambda_{K+1}\ge\lambda^\star$, yielding the rate-improvement truncation
\begin{equation}
K^\star_{\rm rate}=\#\{k:\lambda_k<\lambda^\star\};
\label{eq:D-Kstar}
\end{equation}
beyond it, prethermalization can still reduce prefactors or oscillatory
transients but not the envelope rate. Moreover, since
$r(\lambda_1)\approx\lambda_1/\gamma$ for $\lambda_1\ll\lambda^\star$,
\begin{equation}
S_{\rm ud}(K)\le\frac{\gt}{r(\lambda_1)}\approx\frac{2\lambda^\star}{\lambda_1}:
\label{eq:D-ceiling}
\end{equation}
the Mpemba gain is capped by a purely inertial ceiling that no amount of
digital preprocessing can lift.
 
If instead the hardware is engineered so that $\lambda^\star\le\lambda_1$
(high quality factor), every single-mode envelope decays at the uniform rate
$\gamma/2m$; the slowest covariance envelope then decays at rate $\gamma/m$
and the ensemble-covariance burn-in time,
$t_\varepsilon\simeq(m/\gamma)\ln(C/\varepsilon)$ with $C$ a
threshold-dependent prefactor, is independent of the condition number of $A$
(this does not by itself imply condition-number-independent sampling cost,
since autocorrelation times and finite-time oscillatory transients may still
depend on the spectrum). This apparent advantage comes with a trade-off: the rate
$\gamma/2m$ degrades with the inertia used to reach it. Optimizing over $m$
at fixed $\gamma$, $r(\lambda_1)$ is maximized at critical damping of the
slowest mode, $m^\star=\gamma^2/4\lambda_1$, where
$r(\lambda_1)=2\lambda_1/\gamma$, only a factor of $2$ above the
overdamped rate. The same holds mode by mode: projecting $K$ modes digitally
and retuning $m_K=\gamma^2/4\lambda_{K+1}$ gives
$t_\varepsilon^{\rm opt}(K)\simeq\gamma\ln(C/\varepsilon)/(4\lambda_{K+1})$,
so the speedup scaling $\lambda_{K+1}/\lambda_1$ carries over intact, with
inertia contributing a constant factor $\le2$. Conversely, deep in the
underdamped regime the Mpemba mechanism is neutralized: the real parts of the
spectrum are degenerate and no slow-mode hierarchy remains to exploit.
Inertia and optimized initialization are therefore near-substitutes rather
than multiplicative resources. Two caveats apply, since $\mathbf{F}$ is non-normal:
$\|e^{\mathbf{F}t}\|$ can exhibit transient growth bounded by the conditioning of the
eigenvector matrix, which diverges at critical damping (Jordan block,
$t\,e^{-\gt t}$ prefactors), so in practice mode $K{+}1$ should be slightly
underdamped; and for underdamped modes $\mathbf{\Sigma}_{xx}(t)$ oscillates around its
envelope at frequencies $\Omega_j\pm\Omega_k$, so the covariance estimator
must time-average over a beat period or sample stroboscopically, the
decorrelation time of successive samples growing with the quality factor.

Finally, the RLC capacitance encoding of Ref.~\cite{Melanson2023} corresponds, in
canonical Langevin variables, to the complementary case in which the
programmable matrix appears in the capacitance, or mass, sector rather
than in the stiffness sector. If \(C u_k=c_k u_k\), with scalar inductance
\(L\) and resistance \(R\), each mode obeys
\[
\mu^2+\frac{\mu}{R c_k}+\frac{1}{L c_k}=0 .
\]
Thus the slow mode-dependent branch is the large-\(c_k\) underdamped
branch, with envelope rate \(1/(2R c_k)\), consistent with
\(\tau_{\rm corr}\sim R c_{\max}\). In this encoding the optimized
initialization targets the largest programmable eigenvalues rather than
the smallest ones; apart from this reversal of spectral ordering, the
covariance-matching construction is unchanged.

\suppnote{\updated{Digital cost and the trade-off in $K$}}
\label{app:tradeoff}
The hybrid protocol combines a digital stage (Lanczos, measured in wall-clock
seconds on a classical processor) with an analog stage (relaxation of the
physical hardware). The analog cost is expressed in
dimensionless time or converted to seconds through the hardware time constant
$\tau_c$, the physical duration of one unit of $\mu^{-1}$. Throughout this appendix we set $k_BT=1$.

The Lanczos algorithm resolves the lowest $K$ eigenpairs in
$I(K)\approx\nu K$ iterations, each costing one matrix-vector product
($2d^{2}$ FLOPs for dense $A$, $2\,\mathrm{nnz}(A)$ for sparse $A$), with
reorthogonalization subleading for $K\ll d$. The prefactor $\nu=\mathcal{O}(1)$
when the lower spectral edge is well separated and grows as the relative gaps
$(\lambda_{K+1}-\lambda_{K})/(\lambda_{\max}-\lambda_{\min})$ shrink~\cite{Saad2011}. Writing $F$ for the sustained FLOP rate, the digital
time is thus essentially linear in $K$
\begin{equation}
  T_L(K)\;\approx\;\frac{2\nu\,K\,d^{2}}{F},
  \label{eq:TL}
\end{equation}
so the marginal cost of one additional resolved mode is approximately flat in
$K$ and quadratic in $d$
\begin{equation}
    \Delta T_L\;\equiv\;T_L(K{+}1)-T_L(K)\;\approx\;\frac{2\nu d^{2}}{F}
    \label{eq:dTL}
\end{equation}
and is bounded above by full diagonalization,
$T_L(K)\leq T_L(d)=\mathcal{O}(d^{3})$.

On the analog side, the optimized initialization of Eq. (15) in the main text prethermalizes the
$K$ slowest covariance modes, leaving the error dominated by the slowest
remaining mode $\lambda_{K+1}$; from Eq.~(20) in the main text,
$t_{0}(\epsilon_t,K)\approx L/(2\mu\lambda_{K+1})$ with $L\equiv\ln[E_{0}(K)/\epsilon_t]$
and $E_{0}(K)$ given by Eq.~(35) in the main text. The logarithmic factor $L$ depends only weakly
on $K$ (removing the term $\lambda_{K+1}^{-2}$ from the residual sum shifts $L$
slightly downward, so treating it as constant is conservative). Prethermalizing
mode $K{+}1$ moves the bottleneck rate $\lambda_{K+1}\!\to\!\lambda_{K+2}$,
saving per trajectory
\begin{equation}
  \Delta t_{0}\;\approx\;\frac{L}{2\mu}
  \bigl(\lambda_{K+1}^{-1}-\lambda_{K+2}^{-1}\bigr)
  \;=\;\frac{L}{2\mu}\,\frac{g_{K+1}}{\lambda_{K+1}\lambda_{K+2}},
  \label{eq:dt0}
\end{equation}
with $g_{K+1}\equiv\lambda_{K+2}-\lambda_{K+1}$.
Since thermalization runs once per trajectory, with $P$ trajectories executed in
parallel the total analog saving in seconds is $(N_\mathrm{traj}/P)\,\Delta t_{0}\,\tau_c$.

Increasing $K$ is worthwhile while the marginal digital cost stays below the
marginal analog saving, $\Delta T_L\lesssim(N_\mathrm{traj}/P)\,\Delta t_{0}\,\tau_c$. Using
\eqref{eq:dTL} and \eqref{eq:dt0}, the optimum $K^{\star}$ solves
\begin{equation}
\frac{g_{K+1}}{\lambda_{K+1}\lambda_{K+2}}
\simeq
\frac{4\nu\mu d^2 P}{N_{\rm traj} L F \tau_c}
\equiv
\frac{C_{\rm hw}}{\tau_c},
\quad
C_{\rm hw}\equiv
\frac{4\nu\mu d^2 P}{N_{\rm traj} L F}.
\label{eq:Kstar}
\end{equation}
whose right-hand side is a single dimensionless combination of hardware and
problem parameters. \Cref{eq:Kstar} should be read as an interior optimality
condition. It identifies the value \(K^\star\) at which the extra digital
cost of resolving one more eigenpair is balanced by the analog
thermalization time saved by prethermalizing the corresponding additional
slow mode. The left-hand side is purely spectral and decreases with
$K$, guaranteeing an interior optimum $K^{\star}\ll d$ unless the right-hand
side is vanishingly small. The limit $K=d$ is therefore not a slow regime but a
redundant one: the digital stage has then solved the problem outright, at cost
$\mathcal{O}(d^{3})$ comparable to the target operation itself, and the analog
device reduces to a Gaussian sampler.

\begin{figure}[t]
  \centering
  \subfloat[]{
    \includegraphics[width=0.95\linewidth]{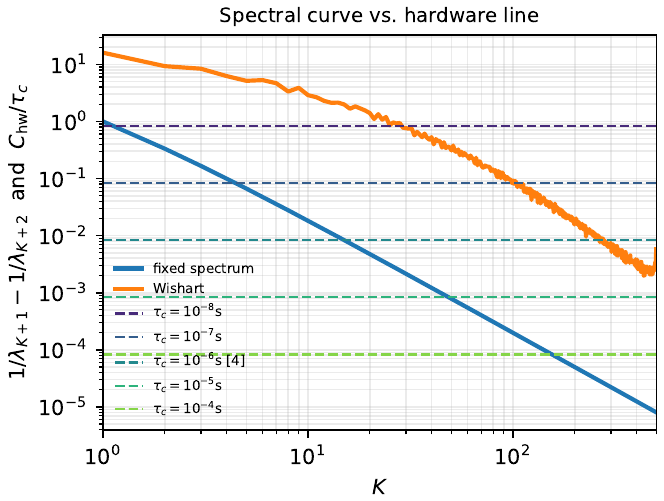}
  }

  \vspace{0.5em}

  \subfloat[]{
    \includegraphics[width=0.95\linewidth]{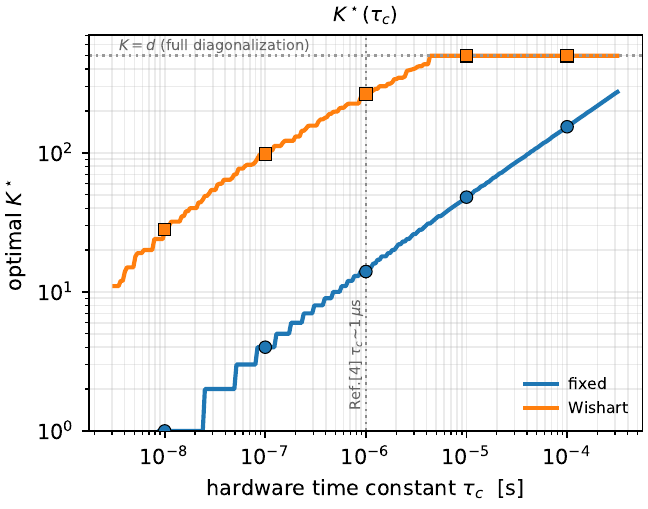}
  }

  \caption{Numerical solution of the optimality condition Eq.~\eqref{eq:Kstar},
  with the hardware time constant $\tau_c$ kept free. (a) the spectral
  marginal saving $1/\lambda_{K+1}-1/\lambda_{K+2}$ versus $K$ for the two
  ensembles (solid), together with the hardware lines $C_{\rm hw}/\tau_c$
  (dashed) for several $\tau_c$; each intersection is the optimum
  $K^\star(\tau_c)$. 
  (b) $K^\star$ read off directly as a function of
  $\tau_c$. Slower hardware (larger $\tau_c$) makes thermalization more
  expensive and pushes $K^\star$ up. The steep $\sim\!1/K^2$ decay of the
  fixed-spectrum curve keeps $K^\star$ modest and weakly $\tau_c$-dependent,
  whereas the crowded Wishart edge gives a shallow curve and a large,
  $\tau_c$-sensitive $K^\star$ that saturates at $K=d$ (dotted) once the
  optimum calls for full spectral resolution and the analog stage becomes
  redundant. Fiducial constants $d=500$, $N_{\rm traj}=10^4$, $\nu=\mu=P=1$,
  $L=12$, $F=10^9\,\mathrm{s}^{-1}$ fix $C_{\rm hw}=4\nu\mu d^2 P/(N_{\rm traj}LF)$;
  changing $\nu$, $L$, or $F$ only rescales the $\tau_c$ axis, so it should be
  read as the combination $C_{\rm hw}/\tau_c$. The value $\tau_c\sim1\,\mu$s
  reported for the hardware of~\cite{Melanson2023} is marked.}
  \label{fig:Kstar}
\end{figure}

Since the left-hand side of Eq.~\eqref{eq:Kstar} is purely spectral, the optimum $K^*$ depends on the class of matrices considered. For the fixed spectrum,
$\lambda_{K+1}=\lambda_{\min}+K\delta$, so the marginal saving decays as
$1/(K^{2}\delta)$ against a flat cost, giving a sharp interior optimum; the
speedup at fixed $K$ is $d$-independent, as the spacing $\delta$ is. For Wishart matrices the lower Marchenko-Pastur edge gives
$\lambda_{K+1}-\lambda_{\min}\sim(K/d)^{2/3}$, so the saving decays only mildly
in $K$ (yielding a broad optimum) but carries a $1/d$ crowding factor.
These behaviors are shown quantitatively in panel b) of Fig.~\ref{fig:Kstar}, where $K^\star(\tau_c)$ is obtained by solving Eq.~\eqref{eq:Kstar} numerically for both ensembles.

\suppnote{\updated{RBF Gram ensemble and regularized inversion}}
\label{app:rbf}
The benchmarks of the main text use ensembles with relatively flat
spectra. In Bayesian machine learning, by contrast, a central
linear-algebra object is the regularized kernel Gram matrix:
Gaussian-process regression, kernel ridge regression, and
Bayesian optimization require solving
$(\mathbf{G}+\sigma^2\mathbb{1})\mathbf{v}=\mathbf{y}$, with
$\mathbf{G}$ the kernel Gram matrix and $\sigma^2$ the
observation-noise variance (or, in kernel ridge regression, the
ridge parameter)~\cite{RasmussenWilliams2006}. This is the ridge
inversion $\mathbf{A}=\mathbf{G}+\epsilon_{\rm reg}\mathbb{1}$
of Eq.~(26) in the main text with
$\epsilon_{\rm reg}=\sigma^2$, and for smooth kernels its
spectrum decays rapidly. We take the
radial-basis-function kernel (rbf)~\cite{RasmussenWilliams2006} on $n=d$
inputs $x_i\sim\mathcal{N}(0,s^2\mathbb{1}_m)$,
\begin{equation}
  \mathbf{A} = \mathbf{G} + \sigma^2\mathbb{1},
  \qquad
  \mathbf{G}_{ij}
  = \exp\!\left(-\lVert x_i-x_j\rVert^2/2\ell^2\right),
  \label{eq:rbf-A}
\end{equation}
which is strictly positive definite.

\begin{figure}[t]
  \centering

  \subfloat[]{%
    \includegraphics[width=0.85\linewidth]{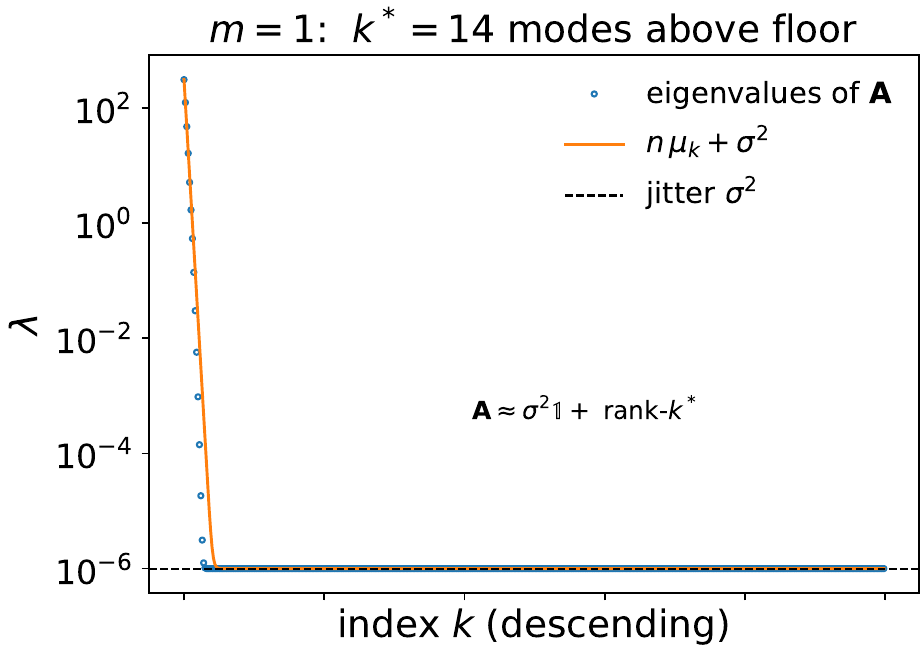}%
  }

  \vspace{-0.3em}

  \subfloat[]{%
    \includegraphics[width=0.85\linewidth]{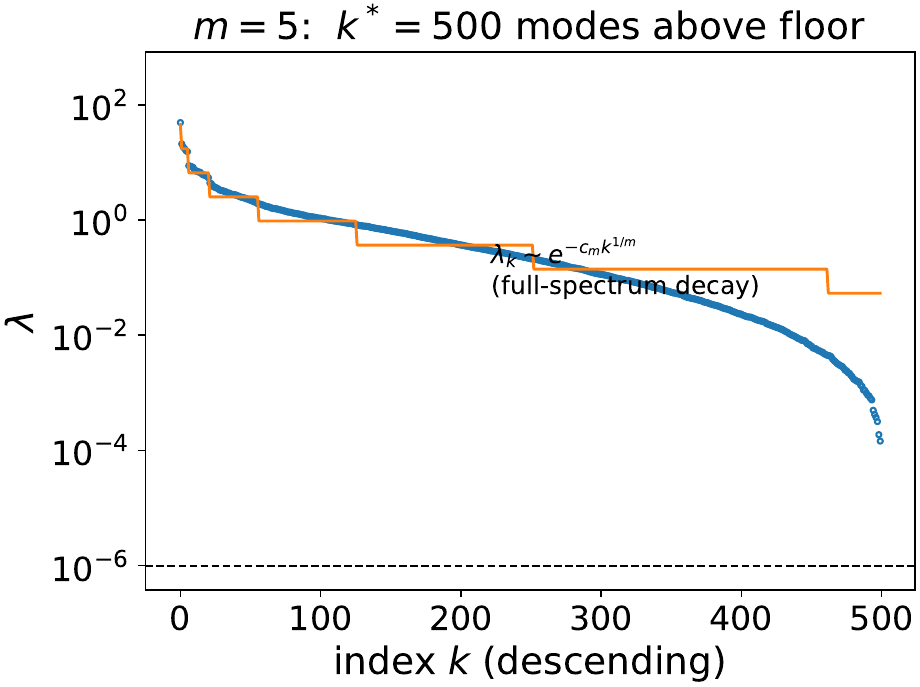}%
  }

  \caption{Spectra of $\mathbf{A}=\mathbf{G}+\sigma^2\mathbb{1}$ for
  $d=500$ inputs from $\mathcal{N}(0,\mathbb{1}_m)$, $\ell=s=1$,
  $\sigma^2=10^{-6}$; the panels differ only in $m$. \emph{a)}
  ($m=1$): only $k^{*}=14$ eigenvalues exceed the jitter, so
  $\mathbf{A}$ is effectively a rank-$k^{*}$ perturbation of
  $\sigma^2\mathbb{1}$ [Eq.~\eqref{eq:lowrank-inv}]. \emph{b)}
  ($m=5$): the law \eqref{eq:stretched} spans the whole spectrum
  ($k^{*}=d$) and the Mpemba protocol attains speedups (see Eq.~(21) in the main text) of  $S(K=25)\approx17$,
  $S(K=50)\approx46$, $S(K=250)\approx1.5\times10^{3}$; this is the
  realization used in Fig.~\ref{fig:rbf-speedup}. Solid lines:
  $n\mu_k+\sigma^2$ from Eq.~\eqref{eq:stretched}.}
  \label{fig:rbf-m-comparison}
\end{figure}

For Gaussian inputs in $m=1$ the operator eigenvalues are geometric,
$\mu^{(1)}_k=\mu_0 B^{k}$, with
$a=1/(4s^2)$, $b=1/(2\ell^2)$, $c=\sqrt{a^2+2ab}$, $A=a+b+c$,
$\mu_0=\sqrt{2a/A}$, $B=b/A\in(0,1)$~\cite{Zhu1998}. As the kernel and
measure factorize, the $m$-dimensional eigenvalues are products of these,
and the sorted sequence obeys a stretched-exponential law
\begin{equation}
  \mu_k \simeq \mu_0^m\exp(-c_m k^{1/m}),
  \qquad c_m = (m!)^{1/m}\ln(1/B),
  \label{eq:stretched}
\end{equation}
with the Gram eigenvalues concentrating around $n\mu_k$ in the bulk. 
This streatched-exponential decay (polynomial for the rougher Mat\'ern kernels)
is standard~\cite{RasmussenWilliams2006,Belkin2018}. Whether $\mathbf{A}$
is worth inverting on thermodynamic hardware is controlled by the effective rank
$k^{*}(\sigma^2)=\#\{k:\lambda_k^{(\mathbf{G})}>\sigma^2\}$: for smooth
kernels on low-dimensional inputs the decay outruns the jitter within a
few dozen modes ($k^{*}\ll d$), while for moderate $m$ one has $k^{*}=d$
at the same $\sigma^2$ (shown in panels a) and b) of~\cref{fig:rbf-m-comparison}, respectively).

When $k^{*}\ll d$, discarding the $d-k^{*}$ Gram modes below the jitter
gives, to the $O(\sigma^2)$ accuracy already conceded by the
regularization, $\mathbf{A}=\sigma^2\mathbb{1}+\sum_{k\le k^{*}}
\lambda_k^{(\mathbf{G})}\mathbf{u}_k\mathbf{u}_k^{T}$: a rank-$k^{*}$
perturbation of the identity. Its inverse is then approximated by
\begin{equation}
  \mathbf{A}^{-1}\approx\frac{1}{\sigma^2}\mathbb{1}
  +\sum_{k=1}^{k^{*}}
  \Big(\tfrac{1}{\lambda_k^{(\mathbf{G})}+\sigma^2}-\tfrac{1}{\sigma^2}\Big)
  \mathbf{u}_k\mathbf{u}_k^{T},
  \label{eq:lowrank-inv}
\end{equation}
requiring only the $k^{*}$ largest eigenpairs of $\mathbf{G}$ (those
Lanczos finds fastest) at $O(d^2 k^{*})$ cost. Here the protocol yields
no speedup ($S_{\rm reg}(K)\approx1$ for $K<d-k^{*}$), but no analog
computation is warranted either, since \eqref{eq:lowrank-inv} already
solves the problem cheaply.

\begin{figure}[t]
  \centering
  \vspace{0.6em}
  \includegraphics[width=\linewidth]{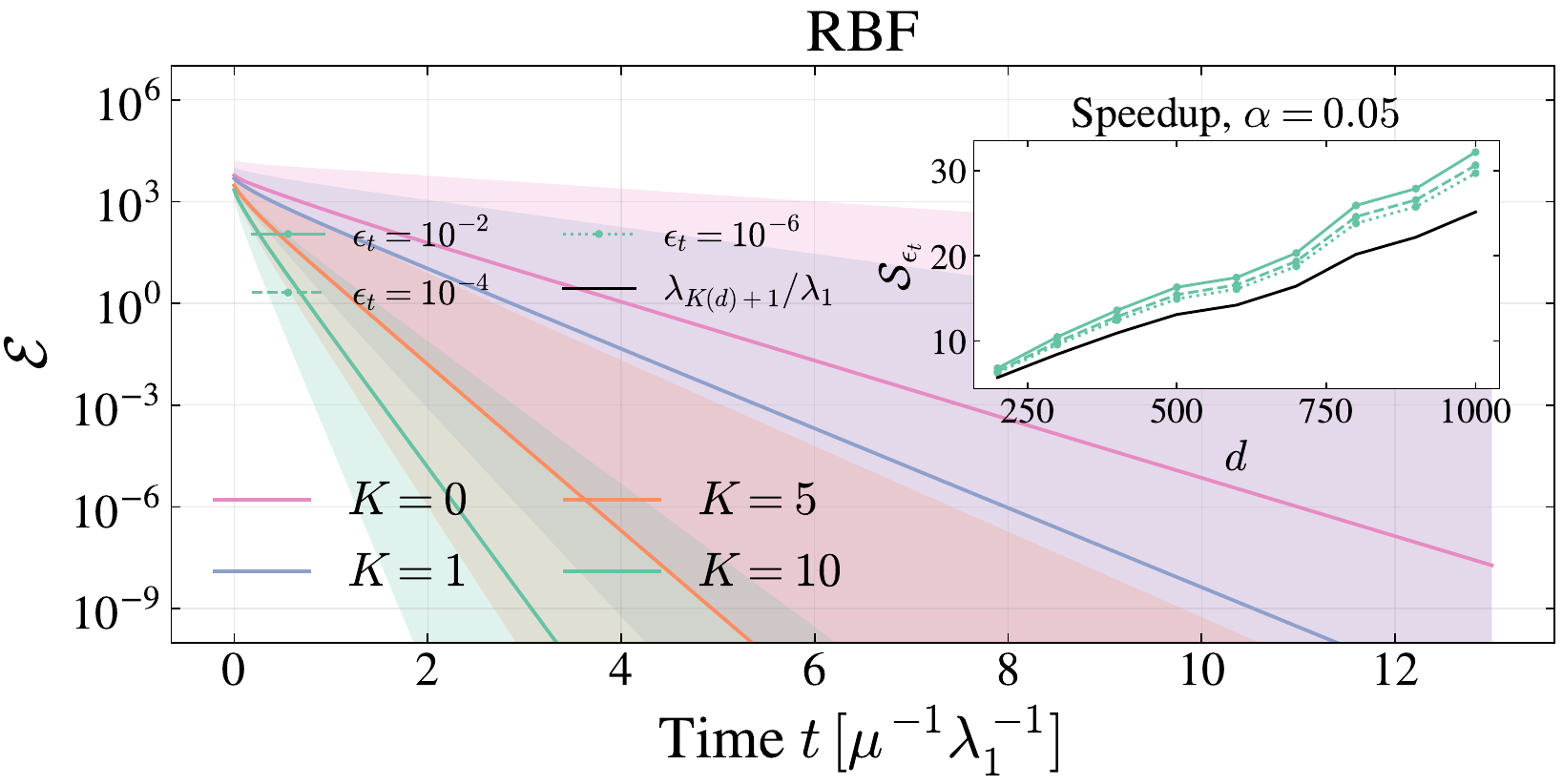}
  \caption{Accelerated inversion of \eqref{eq:rbf-A} with $m=5$,
  $\ell=1$, $\sigma^2=10^{-6}$. \emph{Main:} Frobenius error
  $\mathcal{E}(t)$ at $d=500$ for $K=0,1,5,10$ prethermalized modes
  (medians over $500$ realizations; bands, min--max). The bands are broad
  because $\lambda_1$, which sets the time unit, is the fluctuating tail
  eigenvalue of a stretched-exponential spectrum. \emph{Inset:} speedup
  $S_{\epsilon_t}$ vs.\ $d$ at fixed fraction $\alpha=0.05$ and
  $\epsilon_t\in\{10^{-2},10^{-4},10^{-6}\}$, growing with $d$ and
  tracking $\langle\lambda_{\alpha d+1}/\lambda_1\rangle$ (black). We set
  $\mu=k_BT=1$.}
  \label{fig:rbf-speedup}
\end{figure}

Figure~\ref{fig:rbf-speedup} runs the protocol in the favorable regime
($m=5$, $k^{*}=d$, corresponding to panel b) of~\cref{fig:rbf-m-comparison}). The unassisted dynamics relaxes at $2\mu\lambda_1$;
each prethermalized mode removes a slow direction, so for $K=10$
the error drops below $10^{-9}$ within a few $\mu^{-1}\lambda_1^{-1}$. At fixed $K$ the
speedup decays with $d$ as the jitter absorbs the lowest modes (the
$d$-dependent face of $dS_{\rm reg}/d\epsilon_{\rm reg}\le0$); holding
the fraction $\alpha=K/d$ fixed instead lets the budget grow with the
spectrum, and the speedup rises monotonically over $d\in[200,1000]$,
from $\mathcal{S}\approx6$ to $\approx30$ at $\epsilon_t=10^{-2}$,
tracking $\langle\lambda_{\alpha d+1}/\lambda_1\rangle$. The protocol
thus delivers dimension-robust acceleration on regularized kernel
matrices whenever $k^{*}(\sigma^2)\gtrsim\alpha d$.

\bibliography{Literature}